\documentclass[iop,twocolumn,numberedappendix]{aastex631}
\usepackage{epsfig}
\usepackage{amsmath}
\usepackage{amssymb}
\usepackage{natbib}
\usepackage{hyperref}
\usepackage{graphicx}
\usepackage{mathrsfs}
\usepackage{times}
  \usepackage[flushleft]{threeparttable}
  
\newcommand{\degree}{\ensuremath{^\circ}}

\newcommand{\vect}[1]{\boldsymbol{#1}}

\defcitealias{GLBb}{GLB}
\defcitealias{LBK}{LBK}
\begin{document}
\title{Bottom's Dream and the amplification of filamentary gas structures and stellar spiral arms}
\author[0000-0002-6118-4048]{Sharon E. Meidt}
\affiliation{Sterrenkundig Observatorium, Universiteit Gent, Krijgslaan 281 S9, B-9000 Gent, Belgium}
\author[0000-0002-6118-4048]{Arjen van der Wel}
\affiliation{Sterrenkundig Observatorium, Universiteit Gent, Krijgslaan 281 S9, B-9000 Gent, Belgium}

\begin{abstract}
Theories of spiral structure traditionally separate into tight-winding Lin-Shu spiral density waves and the swing-amplified material patterns of Goldreich \& Lynden-Bell and Julian \& Toomre.  
In this paper we consolidate these two types of spirals into a unified description, treating density waves beyond the tight-winding limit, in the regime of shearing and non-steady open spirals.  This `shearing wave’ scenario novelly captures swing amplification that enables structure formation above conventional Q thresholds.
However, it also highlights the fundamental role of spiral forcing on the amplification process in general, whether the wave is shearing or not.
Thus it captures resonant and non-resonant mode growth through the donkey effect described by Lynden-Bell \& Kalnajs and, critically,  the cessation of growth when donkey behavior is no longer permitted.  
Our calculations predict growth exclusive to trailing spirals above the Jeans length, the prominence of spirals across a range of orientations that increases with decreasing arm multiplicity, and a critical orientation where growth is fastest that is the same for both modes and material patterns. Predicted structures are consistent with highly regular, high- multiplicity gaseous spur features and long filaments spaced close to the Jeans scale in spirals and bars.
Applied to stellar disks, conditions favor low multiplicity ($m<5$) open trailing spirals with pitch angles in the observed range $10^\circ<i_p<50^\circ$.  The results of this work serve as a basis for describing spirals as a unified class of 
transient waves, abundantly stimulated but narrowly selected for growth depending on local conditions. 
\end{abstract}

\section{Introduction}
The origin of spiral structure in gas and stellar disks is a topic with a long, rich history.  The observation that even galaxies with no obvious companions, stellar bars or other source of perturbations can host spiral patterns \citep[e.g.][]{elmegreen82} has led to the understanding that a purely internal process must be responsible in many cases, favoring the notion of self-excited spirals \citep[e.g.][]{linshu,GLBb, JT, SC14}.  From the swing-amplification of wakes around orbiting clusters or clouds \citep{GLBb,JT,toomre81,fuchs1,binney20swing} and global modes \citep{mark74,mark76,bertin89,laubertin} assembled through feedback cycles \citep{mark77,toomre81} to recurrent groove modes \citep{SC14,SC, derijcke} and quasi-static spiral structure \citep{linshu66}, the on-going challenge is to describe how perturbations originate, how they evolve and whether they realistically produce the large-scale trailing spiral structures  observed in grand-design or multi-armed spiral galaxies \citep[see][for a recent review]{sellwoodmasters}.  

In this paper we side-step one of this topic's major complexities 
and assume that all disks are, almost as a rule, already exposed to potential perturbations that originate with both external and internal factors.  This is motivated by the view of galaxies in modern cosmological simulations, where even relatively isolated disks are embedded in and interact with constantly evolving triaxial halos rich in baryonic and dark matter substructures that reflect a history of accretion and merging \citep[e.g.][]{moore99,ferguson02,sancisi08}.  The key to explaining the singular appearance of disk galaxies from this perspective is less a question of how perturbations originate and more an issue of why disks favor the growth of a finite subset of these perturbations.

From a practical point of view, since any periodic potential fluctuation can be expressed as the sum of sinusoids, it represents a rich spectrum of periodic wave perturbations.  
In this work we study the evolution of such perturbations 
to test our hypothesis that the spiral features that dominate a galaxy's appearance have properties set by the disk's own local properties \citep[see also][]{bertin89, linshu}.  This would lead to different characteristic spiral features (radial extent, number of arms) in gaseous disks vs. stellar disks, for example, and would disfavor the formation of non-axisymmetric structure under certain conditions.  It would also suggest that, even if spiral perturbations are each individually transient features \citep[as envisaged by][]{SC14,SC}, long-lasting spiral patterns could develop given that each dominant spiral perturbation would be replaced with a next-generation pattern with similar 
properties.  

Gas disks offer a straightforward illustration of the notion that an array of perturbations present in the underlying disk has consequences for its appearance.   
In the local universe, such gas disks are often exposed to non-axisymmetric stellar bar or spiral patterns in their embedding stellar disks.  
Stellar bars, for example, shape the gas within them into narrow `dust lane' features \citep[e.g.][]{athanassoula92,kim12bars} and set up spirals in the exterior gas \citep{combes85}.  Stellar spiral arms have been extensively studied as a critical factor in the formation of substructures in gaseous spiral arms, including clouds and highly regular spurs and feathers \citep{ko06,shetty,sormani,mandowara}.  
This work contributes another view of how these structures form and evolve, taking insight from new high spatial resolution, high sensitivity JWST/MIRI imaging \citep{leroy23, Sandstrom23,thilker23, meidt23, williams23}.  These images reveal the prevalence of remarkably uniform long filamentary features that extend not only far into the inter arm region between spiral arms, but are also visible alongside dust lanes in bars and in galaxies without prominent underlying spiral arms.  

As we will show in this paper, all of these filamentary gas structures can be described with a singular framework, in which local conditions select a certain set of perturbations for growth.  This basic premise has implications for stellar disks, as well, where 
candidate parent periodic perturbations may be different but no less abundant than in gas disks.  This includes the population of stellar clusters and halo substructures that can collectively perturb the disk \citep{dubinski,nelson12,donghia13}, the halo itself, which can behave like a large embedding bar \citep{franx}, and earlier generations of perturbations that reshape the stellar distribution function in such a way that seeds new perturbations \citep{SC}.  

With this in mind, in this paper we seek predictions for the response of disks to generic $m$-armed spiral potential perturbations in terms of local properties.   We choose an analytical approach that keeps the involved physical factors transparent and can be easily compared to observations and numerical results. 
This work builds on the 
study of axisymmetric stability in the 3D context by \cite{meidt22}.  

The objective here is to obtain an expression that functions like a 3D non-axisymmetric dispersion relation, but in a form flexible enough to capture a snapshot of the stability of generic perturbations that are not necessarily modes or individual shearing wakes but evolving multi-armed patterns.  
Thus our approach deviates somewhat from standard analytical derivations of the non-axisymmetric dispersion relation \citep[i.e.][]{laubertin, GT}.  These were designed to apply in the case that spirals are genuine steady modes of the disk, with a fixed pattern speed $\Omega_p$ at all points along the wave.  They thus implicitly assume that $d \Omega_p/dR=0$ and that the disk is stable or at least near to neutral stability \citep[i.e.][]{laubertin, GT}.     
For our generic scenario, on the other hand, we do not impose a constant pattern speed and explicitly include a growing component for the wave (or complex wave speed).  We also do not place radial boundary conditions to examine the propagation of our shearing spiral waves throughout the disk at this stage. 

The looser stance adopted in this work gives us the powerful facility to place the shearing swing amplification process in the context of the wave framework.  
Part of the underlying motivation to do so is to handle scenarios in which perturbers embedded within the rotating halo or the disk itself collectively produce shearing wave-like perturbations in the disk. Such collective wave perturbations have been found to develop in numerical simulations \citep[i.e.][]{donghia13}.  
Whereas analytical swing amplification calculations effectively treat individual shearing segments separately, 
we work under the assumption that 
collectively produced shearing waves are each candidate spiral patterns with a specific wavelength and (arm) multiplicity or orientation.  We then develop an analytical framework to describe whether they grow or not at their present location, in order to make concrete predictions for the properties of spiral patterns under various circumstances.  

Our approach, described in $\S\S$ \ref{sec:framework}-\ref{sec:additional}, takes the 3D framework used by \cite{meidt22} and adapts it to the study of non-axisymmetry.  Since the perturbations we have in mind are coherent waves in the disk, rather than individual shearing segments (as treated analytically by \citealt{GLBb} and \citealt{JT}) we choose cylindrical coordinates centered on the galaxy rather than shearing coordinates centered on a local overdensity.  

In this context we present an `effective' non-axisymmetric dispersion relation (derived in Appendix \ref{sec:appendix_derivation}) and use it in $\S$~\ref{sec:3dispersionrelation} to identify the existence of two regimes for  structure growth, the second of which is relevant for open shearing spirals after the conventional tight-winding stability threshold is passed.  

To provide some intuition as to how this second `open shearing spiral' regime originates in $\S$\ref{sec:spiralarmframe} we study the perturbed gas response in the spiral arm frame.  In this context we highlight the importance of spiral forcing for factors that favor the growth of arms, specifically radial oscillation (epicyclic motions) and the donkey effect described by \cite{LBK}.  Returning to the non-rotating frame in $\S$~\ref{sec:bottomsdream} we examine the waves that grow most efficiently in the `open shearing spiral' regime, presenting predictions for the critical spiral orientation, or pitch angle, and the growth rate.  We discuss the implications of these predictions for the properties of gas spirals features in spiral arms, bars and in galaxies with no coherent underlying pattern.  We also discuss what these results imply for the structure in stellar disks. 

\section{The stability of non-axisymmetric structures}\label{sec:2}
\subsection{The framework}\label{sec:framework}
The point of comparison of primary interest in the present work is the structure in gas disks, thus we will adopt a framework appropriate for gas in an idealized 3D scenario identical to that used in \cite{meidt22}.  In short, we have an infinitely extended disk (in both the radial and vertical directions) flattened in the vertical direction parallel to the axis of rotation and undergoing non-uniform rotation at a rate $\Omega$ that depends on galactocentric radius.  The gas in this disk is assumed to be approximately isothermal and supported by 3D isotropic non-thermal motions, i.e. with 1D velocity dispersion $\sigma^2=v_s^2+\sigma_{NT}^2$ \citep{chandrasekhar51} that combines the gas sound speed $v_s$ with the non-thermal motions $\sigma_{NT}$. 

With this framework, we obtain an expression for how density perturbations evolve (in the form of a `dispersion relation') 
by combining the continuity equation
 \begin{equation}
\frac{\partial\rho}{\partial t}=\vect{\nabla}\cdot(\rho\vect{v})=0
\end{equation}
with Poisson's equation 
\begin{equation}
\nabla^2\Phi=4\pi G\rho\label{eq:poisson}
\end{equation}
and solutions to the Euler equations of motion for the rotating disk plus a small perturbation, 
\begin{equation}
\frac{\partial\vect{v}}{\partial t}+(\vect{v}\cdot\vect{\nabla})\vect{v} = -\frac{1}{\rho}\vect\nabla p-\vect{\nabla}\Phi. \label{eq:EOM}
\end{equation}
Here $\rho$ is the gas density, $p$ is the thermal plus non-thermal gas pressure \citep[following][]{chandrasekhar51} and the gravitational potential $\Phi$ represents gas self-gravity together with a possible background potential defined by a surrounding distribution of gas, stars and dark matter.  To some extent, this context is also a fair and useful representation of stellar disks, and so the results derived using this framework are also expected to be relevant for describing non-axisymmetry in stellar disks.  A treatment using the collisionless Boltzmann equation \citep[e.g.][]{JT, kalnajs} is beyond the scope of this work.  

To satisfy the linearized perturbed equations of motion in cylindrical coordinates (identical to those in \cite{meidt22}), we adopt local perturbations of the form
\begin{equation}
\Phi_1(R,\phi,z,t)=Re[\mathcal{F}(R,z)e^{i(m\phi-\int\omega dt)}e^{ik R }e^{i k_z z}]  \label{eq:pertex}
\end{equation}
representing the perturbation to self-gravity and 
\begin{equation}
\Phi_{1,b}(R,\phi,z,t)=Re[\mathcal{F}_e(R,z)e^{i(m\phi-\int\omega dt)}e^{ik R }e^{i k_z z}] 
\end{equation}
for a possible external or background potential perturbation, which is assumed to have identical properties and is most useful in the context of this work for visualizing predictions in the absence of self-gravity.  Here $\omega(t)=m\Omega_p+i\omega_i(t)$ is the complex, time-dependent oscillation frequency of the $m$-mode perturbation and the wavenumbers $k$=$2\pi/\lambda_r$ and $k_z$=$2\pi/\lambda_z$ describe the wavelengths of the perturbation in the radial and vertical directions, respectively \citep[e.g.][]{toomre64, GLB, linshu, BT08}.  Variations in the wave pattern speed are taken to be slow compared to the wave growth.

Like all quantities in the equilibrium disk, all wave properties are implicitly functions of galactocentric radius.  The radial variation in $k$ and in $m$ are taken to be weak, however, in order to resemble the shapes of observed spirals.  Thus the distances over which the instantaneous $m$ and $k$ vary are assumed to be considerably larger than the radial wavelength.  A weakening of this assumption might be useful for studying the late-time evolution and propagation of heavily sheared patterns, for example. In the context of the present study, negligible changes in $k$ and $m$ are understood to be a condition for maximal growth at a given $k$ and $m$.  

Since we allow for wave growth and let $d\omega/dR\neq0$, the wave shape (orientation) can also be a function of time (see below).  This variation is assumed to occur through changes in $k$, whereas the spiral multiplicity $m$ remains fixed.  Although in this case the expression we derive from the continuity equation is not a genuine dispersion relation (as discussed at the beginning of $\S$~\ref{sec:approach}), it does provide a view of the instantaneous stability of patterns with a given set properties.  

The density perturbation inherits a form similar to that of the potential perturbation through Poisson's equation (see Appendix \ref{sec:appendix_derivation}),  i.e. $\rho_{1}=\rho_{a}(R,z) e^{i(m\phi-\int\omega dt)}$ where $\rho_{a}(R,z)=\mathcal{R}(R,z)e^{ikR+ik_z z}$.    Solutions to the linearized equations of motion (eq. (\ref{eq:EOM})) also take the form 
\begin{eqnarray}
v_{R,1}&=&v_{R,a}(R,z) e^{i(m\phi-\int\omega dt)},\nonumber\\
v_{\phi,1}&=&v_{\phi,a}(R,z) e^{i(m\phi-\int\omega dt)},\nonumber\\
v_{z,1}&=&v_{z,a}(R,z) e^{i(m\phi-\int\omega dt)}\label{eq:perturbations}
\end{eqnarray}
where $v_{R,a}(R,z)$, $v_{\phi,a}(R,z)$ and $v_{z,a}(R,z)$ are all $\propto e^{ikR+ik_z z}$.

\subsection{The approach: describing swing amplification in the Lin-Shu context}\label{sec:approach}
With this framework we seek an expression for the evolution of the perturbed density $\rho_1$ by combining the continuity equation with Poisson's equation and the solutions to the equations of motion, following an approach close to that used in the tight-winding limit by \cite{linshu}.  We adopt this approach in particular in order to place swing amplification in the context of the conventional Lin-Shu wave framework, merging these two theories of spiral structure into a single theory. In practice, this will give us the ability to compare the thresholds of open spirals with those of their tight-winding counterparts and predict the conditions and properties (wavelength, multiplicity, pitch angle) of spirals that grow to prominence.  

With this design in mind, the approach we use in this section differs from most other treatments of swing amplification, which focus on the force equation \citep[e.g.][]{GLBb,toomre81} to express the reduction of the radial epicyclic frequency in the presence of shearing spirals.  
(Later in $\S$~\ref{sec:spiralarmframe}, though, we place the results of this section in a similar context.) 

\citet[][hereafter GLB]{GLBb} goes beyond the heuristic calculation presented by \cite{toomre81}  and expresses the force equation in terms of the velocity divergence and vorticity, from which point continuity and the conservation of vorticity are used to obtain second order differential equation for $\rho_1$ from the sum of the perturbed forces (eqs. (\ref{eq:radialforce}) and (\ref{eq:azimuthalforce}) in  Appendix \ref{sec:appendix_derivation}).  
That treatment provides a straightforward context for identifying the mechanisms behind swing amplification.   

While an approach like that is certainly possible here, we desire a closer parallel to the \cite{linshu} framework that keeps the factors responsible for swing amplification written in terms of spiral forcing, as opposed to the factor proportional to Oort $B$ identified by \citetalias{GLBb}.  As we will show, this provides an equally concise description of swing amplification and is better suited for identifying the critical orientation where growth is fastest, as opposed to the orientation where the growth rate changes fastest, which it swings through first \citepalias{GLBb}.  

An important and unavoidable caveat in this approach is that the expression for the time evolution of the perturbation we derive is not a genuine dispersion relation since our waves are not steady modes.
This is a consequence of our choice to allow for shearing spirals, for which the frequency $\omega$ has an imaginary component and depends on $R$ and the wavevector varies in time.  Thus, 
we have the effective wavenumber $k_e$:
\begin{equation}
ik_e=i\left(k-\int\frac{\partial\omega}{\partial R}dt\right)\label{eq:ke}
\end{equation}
when calculating the radial gradient in all perturbed quantities.  (The second factor on the right we use later in $\S$~\ref{sec:decay} to identify the saturation of growth.).  

Likewise, the time derivatives of any of the perturbed quantities $q_1$=$\Phi_1$, $v_{r,1}$, or $v_{\phi,1}$ are explicitly
\begin{equation}
i\omega_e\equiv\frac{-\dot q_1}{q_1}=i\left(\omega-R\frac{dk}{dt}\right) \label{eq:tderiv}
\end{equation}
for our scenario of interest, namely a shearing spiral perturbation with a fixed $m$, for which all changes in orientation $\arctan{\gamma}=(kR)/m$ occur through $k=k(t)$.  In the interest of studying fast (instantaneous) growth, we focus on the regime in which we also have $\dot{\rho}_1/\rho_1=i\omega_e$, assuming $\omega_e\gg \dot{k_e}/k_e$ (see Appendix \ref{sec:appendix_derivation}).

In this work, the second term on the right side of eq.~(\ref{eq:tderiv}) becomes $2Am$ for shearing material patterns \citep[see also, e.g.][]{binney20swing}, given that their orientations\footnote{For any arbitrary shearing pattern with speed $\Omega_p$ not necessarily equal to $\Omega$, then $d\tan{\gamma}/dt=-mRd\Omega_p/dR$.}  evolve as $d\tan{\gamma}/dt=2A$ \citep{toomre81,BT08} in terms of Oort $A$
\begin{equation}
A=\frac{-1}{2} R \frac{d\Omega}{dR}.
\end{equation}

With these choices, our expression for $\omega_e(k)$ only admits steady wave solutions in the tight-winding limit \citep[see also][]{binney20swing} and must therefore be used to describe the propagation of wavepackets with care.  
Even so, it remains a powerful tool for identifying the onset of instability; since $2A$ is real, the imaginary part of $\omega_e$ must come from the imaginary part of $\omega$.  We therefore use it specifically to obtain a snapshot of the stability of perturbations at any moment in time.   In what follows we refer to our expression for $\omega_e(k)$ as an effective dispersion relation, specifically with its capacity to diagnose instability in mind.   

\subsection{Recognizing the amplification of spiral structures and the growth of spiral instabilities} 

The main aim of this work is to identify scenarios that lead to spiral growth.  Thus we will use the dispersion relation to seek scenarios in which $\omega$ has a positive imaginary component.  With this condition in mind, the processes of interest should not be regarded as a destabilization of the disk, but rather as the growth of local instabilities, or the formation of structures through local gravitational instability.  We rarely expect our growing spiral waves to destabilize their host disks, given that the growth is almost always transient.  Shearing patterns, for one, undergo growth only up until the time when $k$ swings past $k_J$ \citep[][and see below]{GLBb, JT} and they are stabilized by gas pressure.  Compared to this growth, which we refer to as amplification rather than instability, the growth of genuine wave modes with a fixed pattern speed and radial wavenumber $k$ better qualifies as an instability.  However, we find that this growth is also a local and temporary process (see $\S$~\ref{sec:decay}).  In this light, even though we refer to growing non-axisymmetric structures as instabilities, their presence is not meant to signify 
non-axisymmetric disk instability.    

\subsection{Additional considerations for describing open spirals}\label{sec:additional}
The framework and approach used here is very similar to that used by \cite{meidt22} to obtain the dispersion relation for 3D perturbations (including a vertical component) in the axisymmetric  ($m$=0) limit.  That dispersion relation integrates to a 2D form that resembles the tight-winding dispersion relation derived by \cite{linshu66} and the axisymmetric dispersion relation derived by \citep{kalnajs65}, hereafter referred to together as the Lin-Shu-Kalnajs (LSK) dispersion relation. 

For this work the goal is to fully characterize non-axisymmetric features, and so we deviate from \cite{meidt22} (and Lin, Shu and Kalnajs) by allowing $m>0$ and in particular that $m\sim kR$.  The appendix of \cite{meidt22} presents a narrow version of this calculation, limited in scope to only one of the two growth regimes examined in this work (the non-shearing regime).  
As we will show in what follows, specifically when $kR\gtrsim m$ (beyond the tight-winding limit) the possibility of shearing and the coupling of motions in the radial and azimuthal directions  
(denoted in the linearized equations of motion) gives rise to the second regime of non-axisymmetric instability.  

Besides allowing $kR\gtrsim m$, there are two other relevant constraints on $k$ adopted in this work.  These correspond to either the `short' or `long' wave limits of the general expression for the effective dispersion relation derived in the Appendix.  The short-wave limit is the main focus of this paper and invokes the WKB approximation, which requires that the wave amplitude $\mathcal{R}$ has to vary more slowly with radius than the phase ($k\gg d\ln\mathcal{R}/dR$).  In practice, this is written as $k R_d\gg 1$ with $R_d=(d\ln\rho_0/dR)^{-1}$ given that the wave amplitude must vary at least as fast as the unperturbed disk density ($d\ln\mathcal{R}/dR\gtrsim d\ln\rho_0/dR$), so that the perturbation remains small with respect to $\rho_0$.  A related `short wavelength' constraint assumes $kR\gg 1$, which is well suited for use with the WKB approximation.  Thus, all terms proportional to $1/R$ are negligible unless they specifically involve $m/R$ and $m$ is taken to be large (satisfying $kR\gtrsim m$).  This is equivalent to assuming that all radial variations in the unperturbed disk are much slower than the radial variation of the perturbation.  The treatment of swing amplification made by \cite{GLBb} also makes this assumption, even while handling the scenario $k\rightarrow0$. We will follow the same approach, however acknowledging that the WKB approximation becomes suspect in this scenario and terms of order $1/R$ and $1/R_d$ appearing in the general dispersion relation become relevant.  The preferred treatment in this case would be to take the `long-wave' limit $kR\ll1$, as examined in Appendix \ref{sec:longwave}.  A more detailed study of the growth and propagation of open spirals in this regime is reserved for future work.     

\subsection{Limitations of the present approach}
Although our chosen approach should be sufficient for obtaining insight into the nature of many different types of spirals, there are obvious limitations.  For one, our approach is purely fluid dynamical.  This makes its application to stellar disks limited.  A more appropriate starting point would be to use the Collisionless Boltzmann equation or to adopt a reduction factor to account for the weakening of the self-gravitational force due to phase mixing \citep[i.e.][and references therein]{BT08}.  

Applied to gas disks, our chosen approach also faces some limitations.  Like many of the canonical studies of swing amplification \citep{GLBb} and density waves \citep[e.g.]{linshu, laulinmark, laubertin} in gas disks, we consider only a single disk component with an isothermal equation of state and explicitly neglect factors like magnetic forces, cooling, dissipation, and the injection of thermal and mechanical energy from star formation feedback, for example.  The latter three factors, which here implicitly regulate the equilibrium gas pressure, can impact the support against gravity on different scales \citep{gammie01,romeo10,elmegreen11}.  In magnetically sub-critical gas disks, the support provided by magnetic pressure against gravity shifts the critical collapse scale to the `magneto-Jeans' length, above the Jeans length \citep[e.g.][]{elmegreen87,elmegreen94,kimOstriker01}.  The potential influence of these factors, in addition to the impact of a secondary (stellar) disk on the stability of the gas \citep[e.g.][]{jogsolomon,elmegreen95,jog96,rafikov,romeo11}, should be kept in mind.  

Another important caveat worth noting is that we almost exclusively work in a regime in which the perturbed self-gravity dominates any external or imposed potential perturbation, unlike \cite{JT} and \cite{GT}, for example.  In practice, then, our waves are free solutions to the linearized Euler-Poisson system.  Given that we do not solve an initial value problem in the way that \cite{JT}, \cite{fuchs1} or \cite{binney20swing} do, our calculations do not recover the steady state \cite{JT} ‘wake’ that forms around a heavy perturber on a circular orbit, for example.  Instead, the waves in our description represent the response after the source is removed (i.e. as would be the case when the perturber has a finite lifetime).  This should be adequate for achieving the goals of this work, which is aimed not to describe how perturbations arise but to understand why certain perturbations grow and evolve once present.

\section{The stability of `short wave' open spiral perturbations}\label{sec:shortwavestability}
The effective 3D dispersion relation (eq. [\ref{eq:full3drelation}] derived in Appendix \ref{sec:appendix_derivation} using the framework introduced in $\S$~\ref{sec:framework}) is designed for describing the evolution and stability of open, shearing spiral patterns 
with a range of pitch angles and radial wavelengths, ranging from short WKB spirals with $kR\gg1$ to the longest spiral patterns with $kR\ll1$.  The remainder of this work focuses on predictions in the short wave limit. The long-wave limit (briefly discussed in Appendix \ref{sec:longwave}), like the dispersion relation in general, though, has powerful implications for the propagation of waves and wave packets across the disk.  A comparison between the predictions made with our approach and those described by, e.g., \cite{toomre69}, \cite{mark74}, \cite{toomre77}, and \cite{binney20swing} is reserved for future work.  Here we note that our general non-steady open spiral ``dispersion relation'' is expected to be able to compensate for the limitations of the LSK dispersion relation highlighted by \cite{binney20swing} specifically near corotation.\footnote{Whereas the LSK dispersion relation predicts an evanescent zone around corotation (e.g. Binney \& Tremaine 2008), the dispersion relation in eq. (\ref{eq:full3drelation}) of Appendix \ref{sec:appendix_derivation} allows for transient growing waves to exist throughout the region.}  The general dispersion relation can also be used to 
place new constraints on the group velocity of wave packets propagating away from corotation in the limit $k\rightarrow0$, improving on the view obtained with the LSK dispersion relation (built on the use of the WKB approximation) in this regime. 

\subsection{The `short-wave' open spiral 3D dispersion relation $kR_d\gg 1$, $kR\gg 1$, $kR\gtrsim m$}\label{sec:3dispersionrelation}
The possibility of non-steadiness and radial variation in a spiral perturbation's angular frequency $\omega$ gives the dispersion relation derived in this work (in Appendix \ref{sec:appendix_derivation}) several qualities that differ from most other dispersion relations derived for steady wave modes (discussed briefly in $\S$~\ref{sec:otherRelations}).   Our interest in this work are the qualities that emerge explicitly through a term that is of most importance for non-axisymmetric ($m>0$) perturbations in the short-wave regime, 
with $kR_d\gg 1$, $kR\gg 1$ and $m> 1$.  In these limits, the variation in the background equilibrium disk is essentially negligible \citepalias[also see][]{GLBb}.  

The spirals in this regime are `open',  with orientations that are intermediate to conventional tight-winding spirals (oriented near 10\degree) and the orientations where swing-amplified features switch from leading to trailing (near 90\degree), where the growth rate changes fastest \citepalias{GLBb}.  For convenience, to distinguish this `open spiral' regime from the tight-winding limit $kR\gg m$, in what follows we refer to it as $kR\gtrsim m$, though with the understanding that $m/R$ can also exceed $k$.   

\subsubsection{A cubic dispersion relation}
In the `short-wave' limit, the effective dispersion relation for open spirals takes the compact form
\begin{equation}
(\omega_e-m\Omega)^3=(\omega_e-m\Omega)\omega_0^2+i\omega_{c}^3\label{eq:disprelationMaintext}
\end{equation}
where 
\begin{equation}
\omega_0^2=\kappa^2+\left(k_e^2+\frac{m^2}{R^2}\right)s_0^2+C_z \Delta\label{eq:omega0}
\end{equation}
in terms of the radial epicyclic frequency $\kappa$, $\Delta=\kappa^2-(\omega_e-m\Omega)^2$ and a vertical term $C_z$ (Appendix \ref{sec:appendix_derivation}).  This vertical term is negligible in the regime of greatest interest, namely at the galactic mid-plane where $T_z\rightarrow0$ and in the limit $k_zh\ll1$, which are conditions most favorable to instability, as discussed in \cite{meidt22} (and here in Appendix \ref{sec:appendix_derivation} and in $\S$~\ref{sec:conventionalQ}).

In the second factor on the right hand side, 
\begin{equation}
\omega_{c}^3=\left(\frac{2Ak_em}{R}\right)s_0^2\label{eq:omegacross}
\end{equation}
in the short-wave limit (see Appendix \ref{sec:appendix_derivation}).  We label this with a `c' to denote that it arises with the part of the convective derivative that accounts for changes in radial (azimuthal) flow due to flow in the azimuthal (radial) direction.  
Here 
\begin{equation}
s_0^2=\frac{-4\pi G\rho_0}{k_t^2}\left(1+\frac{\mathcal{F}_e}{\mathcal{F}}\right)+\sigma^2 \label{eq:s0sq}
\end{equation}
where
 \begin{eqnarray}
 k_t^2&=& k_e^2+\frac{m^2}{R^2}+(k_z+iT_z)^2 \nonumber\\
&\approx& k_e^2+\frac{m^2}{R^2}.
 \end{eqnarray}
In what follows we are exclusively interested scenarios in which self-gravity is dominant and $\mathcal{F}_e/\mathcal{F}\ll1$, i.e. at times after the external imposed potential is removed.  However, it is straightforward to modify these predictions to apply with some external driving.  

In this form, the effective dispersion relation in eq. (\ref{eq:disprelationMaintext}) is cubic in $(\omega_e-m\Omega)$ and depicts two routes to instability, through either the first or the second term on the right. The first term (discussed more below) represents the conventional route to $Q$~instability.  
When this term dominates, instability is easily recognizable by the condition $\omega_0^2<0$.  Where $\omega_0^2>0$, however, a second avenue for instability remains open, through the second (constant) imaginary term in eq. (\ref{eq:disprelationMaintext}).  In this case, $(\omega_e-m\Omega)$ is also necessarily imaginary or contains an imaginary component, immediately signifying either the growth or decay of the perturbation depending on the sign of $\omega_{c}$. 

\subsubsection{Relation to swing amplification}\label{sec:relationtoswing}
The term $i\omega_{c}^3$ depicts the process of swing-amplification and it
has a number of qualities that make its impact on structure formation in disks clear.  First, it is absent when $m=0$ and negligible compared to terms of order $k^2$ in the tight-winding limit, which forms the basis of conventional descriptions of instability.  Second, it is negligible whenever $A=0$ in the disk and whenever $\partial \omega_i/\partial R=0$, such as for a genuine  wave mode (see $\S$ \ref{sec:appendix_shear}), emphasizing the essential role of shear in the growth process.  

In other ways, this term functions differently than normally envisioned for swing-amplification and indeed, its present form is unlike that highlighted in the calculations of \citetalias{GLBb} and \cite{toomre81}.
This difference emerges from our chosen approach, as noted in $\S$~\ref{sec:approach}. \citetalias{GLBb} derive a second order differential equation for $\rho_1$ (their eq. 69 or eq. 74) by appealing to vorticity conservation (their eq. 60) to interchange the perturbed vorticity with a term proportional to $B/A\rho_1$, which is the factor they conclude is responsible for swing amplification.  Our approach, on the other hand, seeks a first order differential equation for $\rho_1$ (placing solutions to the equations of motion into the continuity equation) that does not explicitly involve the perturbed vorticity (although vorticity conservation is implicitly assumed) and thus keeps the \citetalias{GLBb} swing amplification term written in terms of the spiral forcing.  
The result, as we show here, is that swing amplification appears in the context of the Lin-Shu approach as an imaginary term proportional to 
$s_0^2$ in the density evolution equation.  It can indeed be shown that a term like this emerges from such an approach even adopting the shearing coordinates of \citetalias{GLBb}.  
In the process of transforming the force equation using vorticity conservation following \citetalias{GLBb}, though, this term gets refactored, leading to the behavior identified by \citetalias{GLBb} and \cite{JT}.    

Setting swing amplification in the Lin-Shu wave context as in eq. (\ref{eq:disprelationMaintext}) provides a new way to visualize how it arises and how it relates to conventional gravitational instability.  These advantages are explored in the following sections, where we highlight the intimate connection between swing amplification and the growth of spiral modes through the donkey effect \citep[][hereafter LBK]{LBK}.  Before examining solutions to the dispersion relation that apply in the open spiral regime, below we first summarize conventional instability in the context of our new effective dispersion relation.

\subsection{Conventional non-axisymmetric tight-winding instability subject to the Q thresholds}\label{sec:conventionalQ}
Under specific conditions, the first term in our new effective dispersion relation (eq.~\ref{eq:disprelationMaintext}) dominates over the second, imaginary term and our dispersion relation becomes an expression of conventional stability.  Disks fall into this regime specifically in the tight-winding limit and otherwise when the pattern is not shearing (as assumed in \cite{meidt22}).  
In this case, 
the mid-plane dispersion relation 
\begin{equation}
(\omega_e-m\Omega)^2=\kappa^2+\left(k_e^2+\frac{m^2}{R^2}\right)4\pi G\rho_0\left(-\frac{1}{k_e^2+\frac{m^2}{R^2}}+\frac{1}{k_J^2}\right)\label{eq:conventionaldisp}
\end{equation} 
is quadratic in $\omega^2$, where $k_J^2=4\pi G\rho_0/\sigma^2$ and for transparency all terms proportional to $m/R$ are left in.  

As shown in \cite{meidt22} \citepalias[see also][]{GLBb}, this can be integrated to yield the 2D Lin-Shu dispersion relation, adopting a redefinition of the potential perturbation as a vertical delta function.  The 2D dispersion relation suggests the 2D stability threshold stability $Q_T= 1$ in terms of the Toomre parameter for gas disks $Q_T=\sigma\kappa/(\pi G\Sigma)$. Remaining in 3D, on the other hand, with eq. (\ref{eq:conventionaldisp}) we identify a slightly modified threshold for instability (see also \cite{meidt22})
\begin{equation}
 Q_M\equiv\frac{\kappa^2}{4\pi G\rho_0}=1-\frac{m^2}{(k_JR)^2}.\label{eq:QMthreshold}  
 \end{equation}
This applies specifically to 3D instability localized to the mid-plane, such that $C_z\rightarrow0$ and $T_z\rightarrow 0$, and assuming that the wave behavior of the perturbation in the vertical direction is negligible ($k_zh\ll  1$).  This set of conditions was shown by \cite{meidt22} to lead to the fastest growing instabilities and is what \cite{GLBb} call the `vertical equilibrium approximation'.  

Eq. (\ref{eq:QMthreshold}) reduces to the axisymmetric threshold $Q_M=1$ calculated by \cite{meidt22} when $m=0$ and is identical to the $m>0$ threshold suggested in the appendix of that paper.   

\subsubsection{2D vs. 3D instability}
In the conventional stability regime, the stability threshold depends on whether structures grow throughout the full disk or only near the mid-plane. 
For the stabilization across the entire disk, from $z=\pm\infty$, the threshold is lower than given by eq. (\ref{eq:QMthreshold}), calculated by GLB as $Q_M\approx0.6$ \citep{GLB}, equivalent to the 2D threshold $Q_T=1$ calculated by \cite{toomre64}. (Stability away from the mid-plane makes it more difficult to destablize the entire disk than the region surrounding the mid-plane; \citealt{meidt22}.) Disks in the conventional stability regime are thus subject to two thresholds on $Q_M$, either at the mid-plane or over the full disk, corresponding to thresholds on the Toomre parameter $Q_T\approx2Q_M^{1/2}\approx2$ (mid-plane) and $Q_T= 1$ (full disk).  

The characteristic scales of structures growing through conventional instability also depend on whether structures grow over the whole disk or only at the mid-plane.  
The minimum scale is the Jeans length $\lambda_J=2\pi/k_J=\sigma^2/\rho$ at the mid-plane and the 2D Jeans length $\lambda_{J,2D}=\sigma^2/(G\Sigma)=\lambda_J^2/(2\pi h)$ when the full disk is subject to fragmentation.  These two scales are different 
in weakly-self-gravitating systems like the gas disks in the local universe, for which $h<\lambda_J$ and $\lambda_J<\lambda_{J,2D}$ \citep{meidt23}.  

\subsection{Non-axisymmetric `short-wave' `open' spirals above conventional Q thresholds: swing-amplification and mode growth}\label{sec:shortwaveinstability} 
Outside of the tight-winding limit, eq. (\ref{eq:disprelationMaintext}) suggests that disks have an expansive capability to support growing non-axisymmetric structures;  this is one of the defining qualities of swing amplification emphasized by \citetalias{GLBb} and \cite{JT} \citep[see also, e.g.][]{jogswing, ko01Qthresh}, who use different approaches to highlight the surprising responsiveness of disks to shearing overdensities above $Q_T=1$.   
In this work we see that as $m$ approaches $kR$ and spirals become more open than their tight-winding counterparts, the combination of spiral forcing and shear allows for growth over an extended range in $k$, well into the range of orientations characteristic of the spirals observed in the gaseous and stellar disks of galaxies.  
In practice, the result is that disks should regularly form instabilities even when satisfying conventional stability thresholds ($Q_M=1$ or roughly $Q_T=2$), as is indeed observed.

\subsubsection{Solutions to the cubic dispersion relation}\label{sec:cubic}
The tendency to growth (instability) in the open spiral $kR\gtrsim m$ regime can be diagnosed by examining solutions to the cubic dispersion relation, 
\begin{equation}
(\omega_e-m\Omega)=\frac{1}{3}\left(\alpha^nD+\frac{3\omega_0^2}{\alpha^nD}\right)\label{eq:3dsoln}
\end{equation}
 where  $n$=0,1,2, $\alpha=(-1\pm \sqrt{-3})/2$,  
 \begin{equation}
D=i\frac{3(-\omega_{c}^3)^{1/3}}{2^{1/3}}\left(1\pm\sqrt{1+\frac{4\omega_0^6}{27\omega_{c}^6}}\right)^{1/3} \label{eq:3dsolnD}
\end{equation}
and we take eq. (\ref{eq:omega0}) for $\omega_0$
in the limit $C_z\rightarrow0$, $T_z\rightarrow0$ and $T_R\rightarrow0$ signifying that instability is localized to the galactic mid-plane.  

For exponential growth, $(\omega_e-m\Omega)$ must have a positive imaginary component.  To identify when this is possible, we can make use of how eq.~(\ref{eq:3dsoln}) simplifies in three regimes. The first of these is the conventional stability regime $Q_M\lesssim1$ ($Q_T\lesssim2$) already discussed in $\S$~\ref{sec:conventionalQ}.  

The second regime spans intermediate $Q_M$ values, specifically those for which the term under the square root in eq. (\ref{eq:3dsolnD}) is close to unity.  In this case, 
the solution to the cubic dispersion relation becomes 
 \begin{equation}
(\omega_e-m\Omega)\approx \alpha^n \frac{i}{2}(-\omega_{c}^3)^{1/3}.\label{eq:interQM}
\end{equation}
It can be shown that this regime extends out to 
$Q_M\approx\sqrt{27}/2\approx2.6$ (or $Q_T\approx3.2$), although the precise value 
depends on $k$, $m/R$ and $k_J$.  The stellar and gaseous disks in nearby galaxies are observed to lie mostly in this regime \citep{boissier03,leroy08,romeo13,westfall14,villanueva21,aditya23}.  

In the limit $Q_M\gg1$ in the third regime, to first order eq. (\ref{eq:3dsoln}) reduces to 
 \begin{equation}
(\omega_e-m\Omega)\approx  \alpha^n\frac{i\sqrt{3}}{3}\frac{(-\omega_{c}^3)}{\kappa^2}.\label{eq:bigQM}
\end{equation}
Growth rates in this case are smaller than in the intermediate regime and proportional to $Q_M^{-1}$.  Any growth when $Q_M\gg1$ thus slows as $Q_M$ increases but only goes to zero in the limit $Q_M\rightarrow\infty$, marking one of the characteristics of non-axisymmetric structure growth above $Q_M>1$; rather than being prevented by traditional thresholds, it is must instead pass specific requirements on the spiral properties $k$ and $m$ encoded in $\omega_{c}$.  

\subsubsection{Conditions for growth I: what we learn from the three roots}\label{sec:growthI}
The solutions to the dispersion relation entail a number of characteristics that have immediate consequences for instability.  Here we first focus on the significance of the three roots of the dispersion relation in the regime $Q_M\gtrsim 1$. 

As highlighted in the Appendix, for `short wave' spirals  
\begin{eqnarray}
(\omega_e-m\Omega)&=&m(\Omega_p-\Omega)+m\frac{\partial (\Omega_p-\Omega)}{\partial R}R\nonumber\\
&+&i\left(\omega_i+\frac{\partial \omega_i}{\partial R}R\right).
\end{eqnarray}
The imaginary term in all three roots implies that each 
depicts instability in one manner or another.  The first $n$=0 root entails the clearest path to instability.  In this case, growth is possible provided that $(-\omega_{c}^3)^{1/3}>0$ and  $m(\Omega_p-\Omega)=0$.  As discussed more below, the former condition is met in differentially rotating disks by trailing spirals above the Jeans length (i.e. with self-gravity dominant over gas pressure, so that $s_0^2$ in eq. (\ref{eq:s0sq}) is negative).  The latter state is achieved by shearing material patterns and spiral modes at corotation.  The growth in these two circumstances is the primary focus of the remainder of this work.  

For the other two roots ($n=1$ and $n=2$), 
\begin{equation}
(\omega_e-m\Omega)=\frac{\pm\sqrt{3}}{4}(-\omega_{c}^3)^{1/3}-\frac{i}{4}(-\omega_{c}^3)^{1/3}\label{eq:tworoots}
\end{equation}
in the regime $Q_M\gtrsim1$.  
Both roots have a non-zero real component, making them applicable to situations either away from corotation or when the spiral has a shear rate that differs from that of the disk. From the imaginary term associated with these roots, we see that spirals in these situations decay under exactly the same conditions that lead to the growth of the $n=0$ root ($(-\omega_{c}^3)^{1/3}>0$).  The expectation is thus that spiral modes decay away from corotation, whether they are inside or outside it.  Likewise, the class of material spirals with $\Omega_p$ even slightly different from $\Omega$ undergo decay.  This might include features in gas disks that evolve subject to forces besides the gravitational forces in the disk (such as feedback, for example).  Until their kinematics are dominated by disk (shear) motion, the dispersion relation implies that such spiral features will decay rather than grow.  

On the other hand, in some scenarios spirals can be expected to have $(-\omega_{c}^3)^{1/3}<0$, in which case the two $n=1$ and $n=2$ roots can also depict growth, as considered more in $\S$~\ref{sec:nonresgrowth}.  This growth is expected to be quite limited for material patterns with a small $\delta=m(\Omega_p-\Omega)$, since these roots imply that the growth rate $\omega_i$=$\pm \delta/\sqrt{3}$.  For spiral modes, the non-resonant growth phase depicted by the $n$=1 and $n=2$ roots is entered very late (after several orbital periods), making it unlikely to be important for gaseous spirals that are more quickly destroyed by other factors, like feedback (see $\S$~\ref{sec:nonresgrowth}).     

\subsubsection{Conditions for growth II: what we learn from the features of $\omega_{c}$ }
The most directly accessible path for `short wave' spirals to become unstable is depicted by the $n=0$ root of the dispersion relation in the regime $Q_M\gtrsim 1$ whenever $(\omega-m\Omega)=0$.  According to eqs. (\ref{eq:3dsoln}) and (\ref{eq:interQM}), the main features of the spirals in this dominant 
regime are determined by the behavior of $\omega_{c}$.  In this light, 
there are two primary features of $\omega_{c}$ that are most important for the types of spirals that are able to grow.  
Firstly, they must be located in a shearing disk.  
Secondly, 
the rate of growth is determined equally by $k$ and $m/R$.  Unlike in the conventional stability regime, the growth of `short wave' spirals above $Q_M=1$ is fastest when $m>0$. It is also fastest for $k>0$ at a given $m$, as is typical for swing amplification, which occurs at its peak at some point after the swing from leading to trailing \citep{GLBb,JT,toomre81}.  (This is slightly modified in the 'long wave regime'; see Appendix \ref{sec:longwave}).  

A detailed look at the growth rates of instabilities is deferred to $\S$~\ref{sec:bottomsdream}, after an examination of forces in the spiral arm frame to obtain a deconstructed view of the factors that lead to swing amplification ($\S$~\ref{sec:spiralarmframe}). 
Here we emphasize that this path to instability applies not only to shearing waves undergoing swing-amplification but also to scenarios when the pattern is not shearing or `swinging' through a range of $k$ as envisioned by \citetalias{GLBb} and \cite{JT}.  This second avenue applies to genuine wave modes at corotation in shearing disks. 

As discussed in $\S$~\ref{sec:spiralarmframe} and $\S$~\ref{sec:bottomsdream}, modes undergoing resonant growth will not be practically different than growing shearing patterns, and the two types of growth attain comparable growth rates.  
According to eq.~(\ref{eq:interQM}) and using the expression for $\omega_{c}$ in eq. (\ref{eq:omegacross}), 
amplifying spirals grow at a rate 
\begin{equation}
\omega_{i}=\left(4\pi G\rho_0\frac{2A k_em}{R}\left(\frac{1}{k_e^2+\frac{m^2}{R^2}}-\frac{1}{k_J^2}\right)\right)^{1/3}\label{eq:omegacrossshear}
\end{equation}
where we have set $\mathcal{F}_e\rightarrow 0$.   

The maximum growth rate attained when $k^2+m^2\ll  k_J^2$ is thus slightly below
\begin{eqnarray}
\omega_{i}^{\rm max}&=&4\pi G\rho_0\left(\frac{(1-\beta)\sqrt{Q_M}}{ \sqrt{2(\beta+1)}}\right)^{1/3}\\
&=&\kappa\left(\frac{1-\beta}{Q_M \sqrt{2(\beta+1)}}\right)^{1/3}
\end{eqnarray}
now writing $2A=(1-\beta)\Omega$ and $\kappa^2=2\Omega^2(1-\beta)$ in terms of the logarithmic derivative of the rotation curve $\beta=d\ln V_c/d\ln R$.  
We thus expect growth to occur with an e-folding time that is on the order of a vertical crossing time $t_{vert}=\pi/\sqrt{4\pi G\rho_0}$, or few epicycles above (since $t_{vert}=Q_M^{1/2}t_{epi}$), although this slows as $Q_M$ increases substantially above unity.  

\subsubsection{A limit to growth: the effects of shear and density enhancement}\label{sec:decay}
The growth in either of the two scenarios discussed above is only ever temporary. 
This is clear when $k_e$ in eq. (\ref{eq:omegacross}) is rewritten using eq.~(\ref{eq:ke}) in terms of $k$ plus a factor $\int\partial\omega/\partial Rdt$, keeping in mind that $k$ can also vary in time.  

Focussing on the secondary regime $Q_M\geq1$ we write 
\begin{eqnarray}
\omega_i^3&=&-i\Gamma4\pi G\rho_0\frac{\omega_ik}{R}\left(\frac{1}{k_e^2+\frac{m^2}{R^2}}-\frac{1}{k_J^2}\right)\nonumber\\
&\times&\left[1+\frac{1}{k}\int\frac{\partial\omega }{\partial R}dt-\frac{i}{k}\int\frac{\Gamma\omega_i }{R}dt\right].\label{eq:genOmega3}
\end{eqnarray}

This expression reduces in two different manners, depending on whether shearing patterns or spiral modes are considered.  For short-wave shearing spirals with $\partial\omega/\partial R=2Am/R$, we have $k_e=k$ since $-i\partial\omega_i/\partial R=2Am/R$ in the limit $kR\gg1$ (Appendix \ref{sec:appendix_shear}) and the term in square brackets in eq. (\ref{eq:genOmega3}) is equal to unity.  Thus, for these shearing spirals

\begin{eqnarray}
\omega_i^3&=&2A 4\pi G\rho_0\frac{km}{R}\left(\frac{1}{k^2+\frac{m^2}{R^2}}-\frac{1}{k_J^2}\right).\label{eq:genOmega3shear}
\end{eqnarray}
As already noted earlier, once $k=k_0+(2Am/R)t$ swings beyond $k_J$, the term in parenthesis becomes negative and growth is replaced by decay.  Thus the shearing nature of gaseous material spirals ensures their transience \citepalias[see also][]{GLBb}.  It should be noted that spirals in stellar disks without pressure only ever undergo a decrease in the growth rate over time and would not enter an equivalent decay phase. 

An increase in pressure over self-gravity, clear in the case of shearing spirals, is also responsible for the eventual cessation of the growth of short wave spiral modes at corotation.  For these modes, the second term in square brackets is zero and initially the third term is also small especially since $kR\gg1$ such that $k_e\approx k$.  But over time, the gravitational force that arises with the growing density amplitude becomes increasingly important (compared to the variation in phase).  According to Poisson's equation (see eq. [\ref{eq:poissonFull}]), the resulting increase in $k_e$ with time leads to a decrease in the perturbation's self-gravity.  Again given that $-i\partial\omega_i/\partial R=2Am/R$ in the short wave limit, then for modes with $\partial\omega/\partial R=i\partial\omega_i/\partial R$, $k_e=k+2Amt/R$ or $k_e\approx 2Amt/R$ at very long times.  Thus, as for shearing spirals, the self-gravity of growing spiral modes decreases over time.  

The time it takes until self-gravity weakens compared to pressure (and mode growth changes into decay) can be estimated from when the term in square brackets in eq. (\ref{eq:genOmega3}) becomes negative. (Note again that the growth of stellar spirals would only be expected to slow, rather than switch to decay.)  This condition implies
\begin{eqnarray}
t_{stop}
&=&\frac{1}{2A \tan{i_p}}\left(\sqrt{\frac{k_J^2}{k^2}-\tan^2{i_p}}-1\right)\\ 
&\approx&\frac{1}{2A }\frac{k_JR}{m}, \label{eq:tstop}
\end{eqnarray}
assuming in the second line that $k_J\gg k$ and $k_J\gg m/R$ such that the mode initially satisfied the condition for instability. 

A rough picture of how much the density grows before growth stops can be obtained by integrating $\omega_i$ up to $t_{\rm stop}$.  A more accurate calculation would require the inclusion of higher-order terms and nonlinear factors neglected in the present linear calculation.  These factors will undoubtedly influence how amplifying features evolve in time and ultimately saturate \citep[e.g.][]{SCnonlin,SCsat}, as will the interplay between multiple disk components \citep{jogswing,wadaspirals,ghoshjog1,ghoshjog2} and non-gravitational factors that can be relevant in gas disks \citep{ko01Qthresh}.  This includes cooling and dissipation \citep{elm89, gammie01,elm11} and magnetic forces \citep[e.g.][]{ko2000,ko02spurs,ko06}. 
Neglecting these additional factors, our linear calculations are nevertheless useful as a reference, giving a preliminary indication of the conditions when growth turns to decay due to weakening self-gravity versus when it might instead be expected to saturate \citep[i.e. in the manner recently discussed by][and discussed further in $\S$~\ref{sec:donkey}]{hamilton}.  

Consider first that the longer $t_{\rm stop}$, the larger the density contrast before growth is suppressed.  From eq. (\ref{eq:tstop}) we can infer that structures closest to the Jeans scale have less time to grow before the onset of decay and, as a result, exhibit smaller density contrasts than structures furthest from the Jeans scale.  Thus, at fixed $k/k_J$, the density contrast when growth turns to decay decreases with increasing spiral pitch angle and, at fixed orientation, longer radial wavelengths can grow larger in amplitude before the onset of decay.  
At the same time, we expect amplifying spirals that rapidly attain a large density contrast to be much more likely to pass into a regime of nonlinear saturation \citep{hamilton} before they ever reach a decay phase.  We thus speculate that 
the growth of spirals is primarily limited by saturation at large wavelengths and small pitch angles (low arm multiplicity).  These are spirals with $t_{\rm stop}$ several times $1/(2A)$.   Spirals with short wavelengths and large pitch angles  (high arm multiplicity), on the other hand, are more likely to stop growing due to weakening self-gravity that shuts off growth before a saturated state is reached.   This conclusion applies most strictly to gaseous spirals, but the slowing growth rates of stellar spirals over time (due an identical weakening of self-gravity) would impart a comparable limit to their density contrasts that could also be reached before they near a saturated state. 

\subsubsection{Non-resonant mode growth}\label{sec:nonresgrowth}
Precisely the same conditions that turn the growth where $m(\Omega-\Omega_p)$=0 to decay 
($\S$~\ref{sec:decay}) also change the decay where $m(\Omega-\Omega_p)\neq0$ into growth (as depicted by the $n$=1 and $n=2$ roots; eq. [\ref{eq:tworoots}], $\S$\ref{sec:growthI}).  
That is, after the time $t_{stop}$ given in eq. (\ref{eq:tstop}) has passed, $s_0^2$ and $(-\omega_{c}^3)^{1/3}$ change sign and the imaginary terms in the $n$=1 and $n=2$ roots correspond to growth.  This suggests that, after an initial period during which spiral arm growth is localized around corotation, amplification eventually takes place preferentially on either side of corotation, where now the wave density undergoes decay.    The mechanism responsible for growth in this scenario will be described in $\S$~\ref{sec:donkey}. Here we note that, in a state of weak self-gravitation, waves can grow by leveraging the fact that they collect material as they sweep across the disk.   
According to the relation between the real and imaginary terms in eq. (\ref{eq:tworoots}) for the  $n$=1 and $n$=2 roots, the growth rate $\omega_i=\pm m(\Omega-\Omega_p)/\sqrt{3}$  is enhanced as the distance from corotation increases.    

It is worth emphasizing that this particular mechanism for non-resonant growth applies to gaseous spirals, many of which may not survive until the non-resonant growth phase is reached, given the influence of star formation and other non-gravitational factors present in these disks.  It is also important to note that the situation in stellar disks is considerably different.  
For stellar spirals there is no pressure term and the only avenue for a change in the sign of $(-\omega_{c}^3)^{1/3}$ is through the reduction factor $\mathcal{F}$, which parameterizes the weakening of self-gravity due to phase mixing.   Taking the expression for $\mathcal{F}$ for the Schwarzschild distribution function in \cite{BT08} (eq. 6.63), we expect sign changes only inside the ILR.  Thus non-resonant growth through the $n=1$ or $n=2$ roots would be possible only in this region of the disk.  

Non-resonant growth can also conceivably be initiated and supported by external factors, as long as these factors reinforce the wave differentially, satisfying $\partial\omega_i/\partial R=i2Am/R$.  In stellar disks, for example, this could be achieved by the addition of newly formed stars at the location of the wave (such as also suggested as a manner to sustain growth by \citealt{SC}).  Star formation rates tend to decrease with radius (together with gas densities), yielding the correct radial behavior to support the growth an underlying mode.  Alternatively, any process tied to the disk dynamical time would be a candidate for initiating the growth of either gaseous or stellar modes.  

\section{Insights into the growth of non-axisymmetric structures: motion in the spiral arm frame}\label{sec:spiralarmframe}
\subsection{Towards conditions for lingering in the spiral arm potential}
One of the characteristics of the $m(\Omega-\Omega_p)=0$ amplification highlighted in the previous section (and discussed in more detail later in $\S$~\ref{sec:bottomsdream}) is that modes and shearing patterns will undergo their fastest growth at the same orientations.  
This is a reflection of the importance of spiral forcing, which is the same at a given $m$ and $k$ whether  the spiral is a material pattern or not.  

To solidify and elucidate this view, in this section we switch to the frame of reference that rotates with the spiral arm perturbation at frequency $\omega=\Omega_p/m$ to obtain a clearer physical picture of the mechanisms that allow for the growth of spiral features.  
We build this picture by examining the motion associated with small displacements from the local potential minimum (density maximum) centered at position ($R_0$,$\phi_0$=0).   Thus we solve the equations of motion 
\begin{equation}
\ddot{\vect r}=-\vect\nabla\Phi-\frac{1}{\rho}\vect\nabla P-2\vect\omega\times\dot{\vect r}
\end{equation}
in terms of $R=R_0+R_1$ and $\phi=\phi_0+\phi_1$ where $R_1$ and $\phi_1$ are the small displacements.  For illustration purposes, we focus on motion in the plane and neglect motion in the vertical direction.  
For our segment of an $m$-armed spiral with wavenumber $k$ we adopt
\begin{equation}
\Phi=\Phi_a \cos(kR_1+m\phi_1+m(\Omega_0-\Omega_p) t).
\end{equation}
This can be visualized as the potential associated with a density wave perturbation
\begin{equation}
\rho=\rho_a \cos(kR_1+m\phi_1+m(\Omega_0-\Omega_p) t), 
\end{equation}
or as only a background wave (in the absence of self-gravity). 
In either case $\Phi_a$ (and/or $\rho_a$)
are assumed to vary negligibly on the scales of interest. 

With this perturbation, to first order the perturbed equations of motion \citep[see also][]{BT08} centered on the spiral segment at radius $R_0$ are written as
\begin{eqnarray}
\ddot{R_1}&+&(\kappa_0^2-4\Omega_0^2)R_1-2R_0\Omega_0\dot{\phi_1}\label{eq:req}\\
&=&k\chi_a \sin(kR_1+m\phi_1+m(\Omega_0-\Omega_p) t)\nonumber
\end{eqnarray}
and 
\begin{equation}
\ddot{\phi_1}+2\Omega_0\frac{\dot{R_1}}{R_0}=\frac{m}{R_0^2}\chi_a \sin(kR_1+m\phi_1+m(\Omega_0-\Omega_p) t).\label{eq:phieq}
\end{equation}
Here
\begin{equation}
\chi_a=\Phi_a+\sigma^2\frac{\rho_a}{\rho_0}.
\end{equation}
and we have assumed that the gas is isothermal with equilibrium density $\rho_0$.
In this scenario, we use that the disk is in centrifugal equilibrium at $R_0$ and $\dot\phi_0=\Omega_0-\Omega_p$ or $\phi_0=(\Omega_0-\Omega_p)t$. 

These two force equations can be combined to examine motion transverse to the spiral arms in the manner discussed by \cite{toomre81} or they can be combined to yield a second order differential equation for $\rho_1$, as derived by \citetalias{GLBb}.  To gain a different perspective on swing amplification, in this section we choose to separately examine the radial and azimuthal accelerations.  The approach is thus closer to the description of motion at Lagrange points in a weak bar potential as treated, e.g. by \cite{pfenniger} and \cite{BT08}.  Compared to the latter, the periodic nature of the spiral perturbation in the radial and azimuthal directions and in time has specific consequences that  
lead to a number of interesting behaviors unique to spiral arms. 

\subsection{Radial motion and the decay of epicyclic oscillation}
We first examine radial motion in the limit that azimuthal excursions $\phi_1$ are small.  We integrate eq. (\ref{eq:phieq}) to obtain an expression for $\phi_1$ that can be substituted into eq. (\ref{eq:req}), i.e.
\begin{eqnarray}
\dot{\phi_1}&=&-2\Omega_0\frac{R_1}{R_0}+c\label{eq:integralphi}\\
&-&\frac{m\chi_a}{R_0^2} \frac{\cos(kR_1+m\phi_1+m(\Omega_0-\Omega_p) t)}{m(\Omega_0-\Omega_p)+k\dot{R}_1+\dot k R_1+m\dot{\phi}_1}\nonumber
\end{eqnarray}
where $c$ is a constant of integration.  Because our interest is not limited to scenarios in which $k\gg m/R_0$ we explicitly leave in the spiral forcing in the second term on the right hand side.

Upon substitution of eq. (\ref{eq:integralphi}) into eq. (\ref{eq:req}), $\ddot{R}_1$ has a term proportional to $\sin(kR_1+m\phi_1+m\Omega_0 t-\omega t)$ and a term proportional to $\cos(kR_1+m\phi_1+m\Omega_0 t-\omega t)$, underlining the importance of both radial and azimuthal spiral arm forcing for radial motion \citepalias[see also][]{LBK}.  

At the potential minimum (density maximum) the cosine term dominates the sine term but the latter has an important  influence for small displacements from the minimum, acting like a force of friction that opposes radial oscillation when gravity dominates over pressure.  The result is that material is able to linger near the potential minimum.  To highlight the motion in this case, we reconsider eqs. (\ref{eq:integralphi}) and (\ref{eq:req}) specifically now also in the limit of small $R_1$, expanding  $\cos(kR_1+m\phi_1+m(\Omega_0 t-\Omega_p)t)$ in eq. (\ref{eq:req}) and $\sin(kR_1+m\phi_1+m(\Omega_0 t-\Omega_p)t)$ in eq. (\ref{eq:integralphi}) 
out to lowest order in $kR_1+m\phi_1$.   
This yields
\begin{equation}
\dot{\phi_1}=-2\Omega_0\frac{R_1}{R_0}-\frac{m\chi_a}{R_0^2}\left(m(\Omega_0-\Omega_p)\right)+k\dot{R}_1+\dot k R_1+m\dot{\phi}_1)^{-1}+c.
\end{equation}
specifically at the spiral arm, using that $\sin(m(\Omega_0-\Omega_p)t)=0$ and $\cos(m(\Omega_0-\Omega_p)t)=1$ at the potential minimum $m(\Omega_0-\Omega_p)t$=(0,$2\pi$).  

There are a few interesting limits appropriate for different scenarios.  The first of these sets $\dot{R}_1$, $\dot{k}$ and $\dot{\phi}_1$ small with respect to $m(\Omega_0-\Omega_p)$ (in $\S$~\ref{sec:awayCR}), while the second takes the reverse limit and applies near corotation (in $\S$~\ref{sec:atCR}).  
\subsubsection{Away from corotation}\label{sec:awayCR}
In the limit that $\dot{R}_1$, $\dot{k}$ and $\dot{\phi}_1$ are small with respect to $m(\Omega_0-\Omega_p)$, eq. (\ref{eq:integralphi}) simplifies to
\begin{equation}
\dot{\phi_1}=-2\Omega_0\frac{R_1}{R_0}-\frac{m\chi_a}{R_0^2m(\Omega_0 -\Omega_p)}\left(1-\frac{k\dot{R}_1+\dot k R_1+m\dot{\phi}_1}{m(\Omega_0 -\Omega_p)}\right)+c\nonumber
\end{equation}
or
\begin{eqnarray}
\dot{\phi_1}&&\left(1-\frac{m^2\chi_a}{R_0^2m^2(\Omega_0-\Omega_p)^2}\right)=\label{eq:phi2}\\
&&-2\Omega_0\frac{R_1}{R_0}+\frac{\chi_a}{R_0^2m(\Omega_0-\Omega_p)^2} \left(k\dot{R_1}+\dot{k}R_1\right)+c\nonumber.
\end{eqnarray}

When the second term in parenthesis on the left hand side of eq. (\ref{eq:phi2}) is small, eq. (\ref{eq:phi2}) can be approximated as 
\begin{eqnarray}
\dot{\phi_1}=&-&2\Omega_0\frac{R_1}{R_0}\left(1+\frac{m^2\chi_a}{R_0^2m^2(\Omega_0-\Omega_p)^2}\right)\\
&+&\frac{\chi_a}{R_0^2m(\Omega_0-\Omega_p)^2}  \left(k\dot{R_1}+\dot{k}R_1\right)+c\nonumber
\end{eqnarray}
expanding the inverse of the term in parenthesis in eq. (\ref{eq:phi2}) to lowest order and keeping only terms to lowest order in $\chi_a$.
 
Upon substituting this expression into eq. (\ref{eq:req}) we see that the radial forcing helps material oscillate around the potential minimum while azimuthal forcing introduces a decay-like term at that location.  In the limit of negligible $\phi_1$, the radial acceleration in eq. (\ref{eq:req}) is rewritten to lowest order in $\chi_a$ as
\begin{eqnarray}
\ddot{R_1}&=&-\kappa_0^2R_1-4\Omega_0^2\frac{m^2\chi_a}{R_0^2m^2(\Omega_0-\Omega_p)^2}R_1+k^2\chi_a R_1\nonumber\\
&+&2\Omega_0R_0\frac{\chi_a}{R_0^2m(\Omega_0-\Omega_p)^2}\left(k\dot{R_1}+\dot{k}R_1\right)\nonumber\\
&+& k\chi_a \sin(m(\Omega_0 -\Omega_p) t)+c\\
\label{eq:reqsmallphir}
\end{eqnarray}
This is the equation of a forced damped oscillator,   
\begin{equation}
\ddot{R_1}=-\omega_{lib}^2R_1
+2\delta_r \dot{R_1}+k\chi_a \sin(m(\Omega_0 -\Omega_p) t)
\end{equation}
where
\begin{equation}
\delta_r=\frac{km\chi_a}{R_0}\frac{\Omega_0}{m(\Omega_0-\Omega_p)^2}, 
\end{equation}
the natural frequency is
\begin{equation}
\omega_{lib}^2=\left(\kappa_0^2-k^2\chi_a+\frac{m^2}{R_0^2}\chi_a\frac{4\Omega_0^2}{m(\Omega_0-\Omega_p)^2}\left(1-\frac{R_0\dot{k}}{2\Omega_0}\right)\right)\label{eq:wlib}
\end{equation}
and the driving frequency is $m(\Omega_0 -\Omega_p)$.  As is convention, constants $c$ are absorbed by a shift $R_1\rightarrow R_1+c$ (Binney \& Tremaine 1987).  

According to the equation of motion in eq. (\ref{eq:reqsmallphir}), material at the spiral arm ($\phi_0= 0)$ undergoes radial oscillation at frequency $\omega_{lib}$ over small distances in the potential.
In the absence of spiral forcing ($k\rightarrow0$ and $m/R\rightarrow0$), this radial oscillation occurs at characteristic frequency $\kappa_0$.  Radial forcing will tend to increase the oscillation speed (i.e. when self-gravity dominates over pressure, so that the second term in the expression for $\omega_{lib}$ in eq. (\ref{eq:wlib}) is positive). But the oscillation speed gets reduced with the addition of azimuthal forcing (as denoted by the third negative term in eq. [\ref{eq:wlib}]).  This reduction can become substantial as $m/R$ approaches $k$.  

Azimuthal forcing acts in further opposition to radial oscillation by introducing a decay term in eq. (\ref{eq:reqsmallphir}), which denotes decay whenever $\delta_r<0$ .  This term originates with the Coriolis force that crosses azimuthal motion into the radial direction and acts in the opposite sense to the restoring force supplied by gravity.  The result is a decay of oscillation exclusively for trailing spirals above the Jeans length, whenever self-gravity dominates gas pressure.  

\subsubsection{At corotation}\label{sec:atCR}
When the time evolution of the perturbation is by far dominated by $k\dot{R}_1+\dot k R_1$, such as for shearing material patterns or modes at corotation, eq. (\ref{eq:integralphi}) becomes
\begin{eqnarray}
&&\dot{\phi_1}\left(k\dot{R}_1+\dot k R_1+2\Omega m\frac{R_1}{R_0}\right)+m\dot{\phi}_1^2\nonumber\\
&=&\frac{m}{R_0^2}\chi_a-2\Omega\frac{R_1}{R_0}\left(k\dot{R}_1+\dot k R_1\right)+c, 
\end{eqnarray}
which can be solved in different limiting scenarios.  We will consider two limits, in which either the Coriolis force or the perturbed gravitational force dominates, i.e. the limit $(k\dot{R}_1+\dot k R_1+2\Omega mR_1/R_0)^2\gg \sqrt{m/R_0^2\chi_a}(k\dot{R}_1+\dot k R_1)$ or its opposite.  The first case yields
\begin{equation}
\dot{\phi_1}\approx-\left(\frac{1}{m}\right)\left(k\dot{R}_1+\dot k R_1+2\Omega m\frac{R_1}{R_0}\right)+c.\label{eq:phisoln}
\end{equation}
Upon substituting eq. (\ref{eq:phisoln}) into eq. (\ref{eq:req}) we find 
\begin{equation}
\ddot{R_1}=-\left(\kappa_0^2-k^2\chi_a+\frac{2\Omega_0R_0 \dot{k}}{m}\right)R_1-\frac{2\Omega_0R_0 }{m}k\dot{R}_1+c
\end{equation}
again inspecting the gravitational force on the righthand side in the limit of negligible $\phi_1$.  

Now setting $d\tan \gamma/dt=2A$ for swinging, shearing material arms, the second term in parentheses is $2\Omega2A=4\Omega(\Omega+B)$.  Since $B<0$, and specifically $4B\Omega=-\kappa^2$, the equation of motion becomes   
\begin{equation}
\ddot{R_1}=-\left(4\Omega_0^2-k^2\chi_a\right)R_1-2\Omega_0\tan{\gamma}\dot{R}_1+c
\end{equation}
once again describing decaying radial oscillation.  In contrast to the eq. (\ref{eq:reqsmallphir}), here the azimuthal forcing helps slightly boost the oscillation frequency above $\kappa_0^2$ (since $4\Omega^2>\kappa^2$), but as before, it also acts in opposition to radial oscillation through the introduction of a decay term.  The decay in this case occurs whether or not gravity exceeds pressure as long as the spiral is trailing ($\gamma>0$).  

The suppression of radial oscillation is even more extreme in the opposite limit $(k\dot{R}_1+\dot k R_1+2\Omega mR_1/R_0)^2\gg \sqrt{m/R_0^2\chi_a}(k\dot{R}_1+\dot k R_1)$, when the azimuthal motion is dominated by the gravitational force and 
\begin{equation}
\dot{\phi_1}\approx\sqrt{\frac{\chi_a}{R_0^2}}.  
\end{equation}
In this case, the radial equation of motion (eq. [\ref{eq:req}]) becomes
\begin{equation}
\ddot{R_1}=-\left(\kappa_0^2-4\Omega_0^2-k^2\chi_a\right)R_1-2\Omega_0R_0\sqrt{\frac{\chi_a}{R_0^2}}+c.
\end{equation}
Since generally $4\Omega_0^2>\kappa^2$, this can signify the complete suppression of radial epicyclic oscillation, i.e. as long as $4\Omega_0^2-\kappa^2\sim2\Omega^2$ stays larger than $k^2\chi_a$.  

\subsection{Azimuthal motion and the donkey effect}\label{sec:donkey}
Inspection of azimuthal motion in the limit of small $R_1$ provides an even more powerful view of the mechanisms that allow  spiral forcing to help waves grow.  
We follow \cite{BT08} (section 3.3.3.b) and start in a scenario where the radial equation of motion is written 
\begin{eqnarray}
(\kappa_0^2&-&4\Omega_0^2)R_1-2R_0\Omega_0\dot{\phi_1}\\
&=&k\chi_a \sin(kR_1+m\phi_1+m\phi_0)\nonumber\\
&\approx& k\chi_a \left[(kR_1)\cos(m\phi_1+m\phi_0)+\sin(m\phi_1+m\phi_0)\right]\nonumber
\end{eqnarray}
allowing that the radial force on the right-hand side may be non-negligible for open spirals and, in the second line, expanding to lowest order in $R_1$.  

Now, if we restrict our view and look also in the limit of small $\phi_1$ then \begin{eqnarray}
(\kappa_0^2-4\Omega_0^2)R_1&-&2R_0\Omega_0\dot{\phi_1}\label{eq:phi1dot}\\
&=&k\chi_a \left[(kR_1+m\phi_1)\cos(m\phi_0)+\sin(m\phi_0)\right].\nonumber
\end{eqnarray}
For transparency, in what follows we will retain the factor $\cos{(m\phi_0)}$ but set $\sin{(m\phi_0)}$ to zero, given our exclusive focus towards the potential minima and maxima.  

Taking the time derivative of eq.~(\ref{eq:phi1dot}) implies 
\begin{eqnarray}
\Delta_s\dot{R}_1&=&2R_0\Omega_0\ddot{\phi_1}\\
&+&k\chi_a(\dot{k}R_1+m\dot{\phi_1})\cos(m\phi_0)+c\nonumber \label{eq:phi1dotderiv}
\end{eqnarray}
where
\begin{equation}
\Delta_s=(\kappa_0^2-4\Omega_0^2-k^2\chi_a\cos(m\phi_0))
\end{equation}
and all constants are subsumed into a single factor $c$.   

We use this expression to replace $R_1$ in the azimuthal equation of motion (eq. \ref{eq:phieq}) in three different regimes.  In the first of these, we take the limit $m/R_0\gg k$ and find
\begin{eqnarray}
\ddot\phi_1&=&\left(\frac{\Delta_s}{\Delta_s+4\Omega^2}\right)\frac{m^2}{R_0^2}\chi_a \phi_1\cos{(m\phi_0)}\nonumber\\
&+&\frac{2\Omega_0km\chi_a}{R_0(\Delta_s+4\Omega^2)}\dot{\phi}_1\cos{(m\phi_0)}+c\label{eq:ksmall}
\end{eqnarray}
In the second regime we let $k\sim m/R_0$ and find
\begin{eqnarray}
\ddot\phi_1&=&\left(\frac{\Delta_s+k^2\chi_a}{\Delta_s+4\Omega^2}\right)\frac{m^2}{R_0^2}\chi_a \phi_1\cos{(m\phi_0)}\nonumber\\
&-&\frac{2\Omega_0m}{R_0}\frac{k^2\chi_a \dot{k}}{\Delta_s(\Delta_s+4\Omega_0^2)}\chi_a \phi_1\cos{(m\phi_0)}\nonumber\\
&-&\frac{4\Omega_0^2k\chi_a\dot{k}}{\Delta_s(\Delta_s+4\Omega_0^2)}\dot{\phi}_1\cos{(m\phi_0)}+c,
\label{eq:kmed}
\end{eqnarray}
or, taking $\dot{k}=2Am/R_0$,
\begin{eqnarray}
\ddot\phi_1&=&\left(\frac{\Delta_s+k^2\chi_a}{\Delta_s+4\Omega^2}\right)\frac{m^2}{R_0^2}\chi_a \phi_1\cos{(m\phi_0)}\nonumber\\
&+&\frac{2\Omega_0km\chi_a}{R_0(\Delta_s+4\Omega^2)}\dot{\phi}_1\cos{(m\phi_0)}+c\label{eq:kmed2}
\end{eqnarray}
Finally, in the limit $k\gg m/R_0$ we find
\begin{eqnarray}
\ddot\phi_1&=&\frac{-2\Omega_0km\chi_a}{R_0(\Delta_s+4\Omega^2)}\dot{\phi}_1\cos{(m\phi_0)}\left(1+\frac{\dot{k}}{\Delta_s}\frac{2\Omega_0R_0}{m}\right)\nonumber\\
&-&\frac{2\Omega_0m}{R_0\Delta_s}\frac{(k\chi_a)^2\dot{k}}{(\Delta_s+4\Omega^2)}\phi_1\cos{(m\phi_0)}\label{eq:klarge}
\end{eqnarray}
which implies
\begin{eqnarray}
\ddot\phi_1&\approx&\left(\frac{2\Omega_0km\chi_a}{R_0(\Delta_s+4\Omega^2)}\right)^2\phi_1\cos{(m\phi_0)}\label{eq:klarge2}
\end{eqnarray}
when $\dot{k}\rightarrow0$
or
\begin{eqnarray}
\ddot\phi_1&=&\left(\frac{k^2\chi_a}{\Delta_s+4\Omega^2}\right)\frac{m^2}{R_0^2}\chi_a \phi_1\cos{(m\phi_0)}\label{eq:klarge3}
\end{eqnarray}
when $\dot{k}=2Am/R_0$.  

All three of these equations depict the donkey effect, or inverse Landau damping, described by Lynden-Bell \& Kalnajs (1972).  For reference, in the absence of radial forcing the equations of motion (eqs. [\ref{eq:req}] and [\ref{eq:phieq}]) imply
\begin{equation}
\ddot{\phi_1}=\left(\frac{\kappa^2-4\Omega_0^2}{\kappa^2}\right)\frac{m^2}{R_0^2}\chi_a \phi_1\cos{(m\phi_0)}, \label{eq:refdonkey}
\end{equation}
The change in angular momentum upon approaching a spiral arm shifts orbiting material to smaller galactocentric radius where it must speed up, given that in most galaxy disks $d\Omega/dr<0$ and $\kappa^2<4\Omega^2$.  As a consequence, material can be either trapped or preferentially driven from certain locations in the potential, as discussed by \cite{BT08} in the context of bars \citep[see also][]{pfenniger}.  Specifically at maxima of the potential $\cos{(m\phi_0)}=-1$, such as at Lagrange points L4 and L5 at corotation in bars, eq. (\ref{eq:refdonkey}) becomes the equation of motion for a harmonic oscillator, describing the libration of material around those locations.  At potential minima, in contrast, the donkey effect leads to the opposite behavior and pushes material away from the minimum, for example making the L1 and L2 points unstable saddle points in the bar potential.  

Likewise, at the corotation circle in spiral potentials, eq. (\ref{eq:refdonkey}) predicts that material will librate in the middle of the interarm region between spiral arms and move to avoid the spiral minima.  This behavior is the small excursion (small $\phi_1$ and small $R_1$) limit of the nonlinear phenomenon of horseshoe orbits \citep[e.g.]{sellwoodbinney} considered by \cite{chibaSch} and \cite{hamilton}.  \footnote{The pendulum equation satisfied by $\phi_1$ that describes motion on horseshoe orbits \citep{chiba23,hamilton} becomes eq. (\ref{eq:refdonkey}) near the center of libration.} The result is that material entering the arm piles up on the 'downhill' side of the spiral, which \citetalias{LBK} showed is the key process that allows material to exchange energy and angular momentum with the wave at corotation, ultimately leading to wave growth\footnote{When the wave represents an external potential perturbation, the result is the excitation of short trailing waves  \citep{GT}.}.  

The donkey behavior depicted in the limit $k\gg m/R$ is precisely the same, even if it appears in a slightly different form: material arriving at the spiral piles up on the upstream side of the arm.  This allows us to clearly see that resonant exchanges through the donkey effect are the same whether the wave is leading or trailing (eq. [\ref{eq:klarge2}]), but only trailing spirals are able to grow  (eq. [\ref{eq:klarge}]; see also \citetalias{LBK}).  The characteristic timescale in this scenario is more closely parallel to $1/\omega_{c}$ derived in a similar limit in the context of the continuity equation.  

\subsubsection{An end to donkey behavior}
The effectiveness of growth through the donkey effect at corotation can be assessed with the libration frequency (i.e. the first term on the right of eq. [\ref{eq:kmed}]), which increases as the perturbed density grows.  Eventually, however, spiral perturbations are expected to follow one of two possible paths that lead to a suspension of donkey behavior and growth.  In the first case, growth saturates as a result of the nonlinear motions that lead to horseshoe orbits \citep{hamilton}.  The motions in the nonlinear regime cause the density response at the arm to disperse \citep{SCsat}, shutting off the donkey effect, and (as shown for stars) leading to phase mixing near the corotation resonance \citep{hamilton}.  Although the small-angle treatment here does not describe the actual saturation process, 
it provides a useful depiction of the conditions when the nonlinear regime is entered, 
which is roughly when the libration frequency is much faster than growth rate.  That is, growth remains possible as long as  
\begin{equation}
\frac{2\Omega^2 km\chi_a  }{R_0\left(\kappa^2-k^2\chi_a \right)}<\omega_i^2 \label{eq:satcondition}
\end{equation}
in the limit $k\sim m/R_0$ (using the libration frequency in eq. [\ref{eq:kmed}] in the case $\dot{k}=0$).  This condition, perhaps not surprisingly, fails to be met when the perturbed density grows comparable to the equilibrium density.  For waves with $k\gg k_J$, eq. (\ref{eq:satcondition}) leads to roughly 
\begin{equation}
\frac{2\Omega 4\pi G\rho_a  }{\left(\kappa^2+4\pi G\rho_a \right)}<(2\Omega 4\pi G\rho_0)^{2/3} 
\end{equation}
in the regime $Q_M\gtrsim 1$
or
\begin{equation}
\frac{\rho_1}{\rho_0} <\frac{Q_M}{Q_M^{1/3}-1}.
\end{equation}
We refer the reader to \cite{hamilton} for a fully nonlinear calculation of the spiral saturation amplitude, noting here that we can expect the donkey behavior to cease being important when the density contrast exceeds $\sim$ 0.1.  

The second avenue to suspend donkey behavior is present at corotation (or for shearing patterns with $\Omega_p=\Omega$) in the linear regime.  This is when self-gravity weakens relative to gas pressure, either due to changes in the spiral shape (in the case of material patterns) or to the growth of the density perturbation, as described in $\S$~\ref{sec:decay}.   The same condition that leads to a decay of growth leads to a sign flip to the right hand sides of eqs. [\ref{eq:ksmall}] - [\ref{eq:klarge}].  The result is the libration of material around the potential minimum, rather than the donkey pile-up away from the spiral potential minimum.  

\subsubsection{Non-resonant growth through the donkey effect}
Donkey behavior is also possible even away from corotation, thereby leading to non-resonant wave growth.  The importance of non-resonant exchanges for wave growth in a regime of fast growth was already noted by \citetalias{LBK} alongside the main result of their work, which is that {\it steady} waves produce no lasting secular changes except at resonances \citepalias{LBK}.   Our calculations here and in $\S$~\ref{sec:shortwavestability} suggest two possibilities for growth away from resonance.  First, donkey behavior could occur temporarily at any location along the arm while $\phi_0=(\Omega-\Omega_p)t$ remains small (near to the potential minimum).  The slower the exit from the spiral arm compared to the timescale to shift `downhill', the more important non-resonant donkey behavior can become.  In this sense, material patterns would have a clear advantage, but so might wave modes that are already growing \citepalias{LBK}.  
In this case, writing $m\phi_0=i\omega_i t$ at the spiral minimum, then for non-resonant exchanges to be important we must have 
\begin{equation}
\left(\frac{2\Omega^2 km\chi_a  e^{\int\omega_i dt}}{R\left(\kappa^2-k^2\chi_a e^{\int\omega_i dt}\right)}\right)^{1/2}>m(\Omega-\Omega_p)\label{eq:fastregime}
\end{equation}
in the limit $k\sim m/R$.
The condition given in eq. (\ref{eq:fastregime}) is equivalent to what \cite{TW84} call the `fast  regime’ \citep[see also][]{weinberg04, weinbergKatz07a,weinbergKatz07b, chiba23}.  Since the amplification is possible only temporarily while $\phi_0=(\Omega-\Omega_p)t$ is small, this form of non-resonant growth may require multiple spiral arm passages to build up, perhaps resulting in growth at a rate proportional to $m(\Omega-\Omega_p)$ (as required by the second two roots to the dispersion relation; $\S$ \ref{sec:nonresgrowth}).  

A second avenue for the growth of gaseous spirals away from corotation involves a slightly different form of donkey-like behavior.  As portrayed by the libration equations (eqs. [\ref{eq:ksmall}], [\ref{eq:kmed}] and [\ref{eq:klarge}]), when pressure forces rather than gravitational forces dominate at the spiral arm, material oscillates around the potential minimum, spending the bulk of its time away from the minimum itself.  (This behavior is responsible for one of the paths for shutting down donkey behavior at corotation discussed in the previous section.) For wave growth to be possible under these circumstances, a factor similar to donkey behavior must be present to counter this motion.  This role can be played by the passage of the arm across the disk, which acts to continuously push material toward the minimum, at least temporarily countering the pressure forces.  The faster the inward flow of material, set by the rate $m(\Omega-\Omega_p)$, the more effective (and fast) the growth.  In this scenario, the sense of the arm (leading or trailing) required for growth is unimportant for the actual growth rate, but trailing spirals will reach the non-resonant growth phase earlier.  

An end to the growth in this scenario would be possible when $-k^2\chi_a$ exceeds $\kappa^2$, flipping back the sign of the denominator of the libration frequency.  However, given that gas spirals would only reach this phase of growth after many orbital periods (see $\S$~\ref{sec:nonresgrowth}), we find it likely that they will be first destroyed by other factors (like feedback, dissipation or interactions with other overdensities not accounted for in the present calculations). 

\subsection{Summary}
Spiral arm forcing in the regime of open spirals produces several behaviors that favor spiral growth at corotation. These behaviors are not present for either tightly-wound spirals or segments oriented at 90$^\circ$ because they capitalize on comparable forcing in both the azimuthal and radial directions. We find that when gravity dominates over pressure, radial oscillation is slowed and can even be suppressed, bringing material closer to the potential minimum.  Meanwhile, in the azimuthal direction the combination of gravity and rotation allow for the `donkey behavior' that leads to wave growth \citep{LBK}.  This also applies to shearing material patterns, which behave just like modes undergoing resonant growth.  In both cases, growth can be attributed to the donkey effect.  

The growth made possible by donkey behavior is only ever temporary, lasting either until the onset of nonlinear saturation \citep{SC, hamilton} or, for gaseous spirals, until self-gravity weakens relative to pressure.  According to the equations of motion, since pressure forces act reversely to gravitational forces, donkey behavior ceases when pressure dominates.   

At the same time, the weak self-gravity, pressure-dominated conditions that disfavor growth at corotation might be the ideal conditions for supporting wave growth at locations away from resonance.  This non-resonant growth relies on the constant supply of material to the spiral potential as the wave passes across the disk, which has a similar effect as donkey behavior, countering motion that would otherwise depopulate the arm.  Given that spirals must persist for several orbital periods to enter this regime, however, the likelihood of this scenario is almost certainly limited by the reality of conditions in gas disks, where factors like dissipation and star formation feedback may more quickly destroy spiral features.   

\section{The characteristics of spiral features growing through donkey behavior (`Bottom's Dream')}\label{sec:bottomsdream}
In the previous section we highlighted how material orbiting around in the potential of a disk with an open spiral behaves like Lynden-Bell's donkey, slowing as it approaches the spiral minimum the more it is pulled radially by the spiral.
Taking the motion of neighboring parcels of material together, for continuity to be satisfied in the context of a differentially rotating disk, the result is a build-up of gas density that reinforces the wave, most effectively when the angular speed of the wave (real or imaginary) varies with radius similarly to the disk itself.  Eventually, the response to the growing wave changes and donkey behavior ceases, causing the wave growth to cease.\footnote{The response of material orbiting in the presence of a spiral wave can be compared to the behavior of Shakespeare's Nick Bottom (from the comedy `A Midsummer Night's Dream').  After being transformed with the head of an ass by the forest sprite Puck and then welcomed to the side of the fairy Queen Titania, Bottom awakens from what he gathers is a dream and commits to recording these events as a ballad called 'Bottom's Dream'. }    

\begin{figure*}[t]
\begin{center}
\begin{tabular}{cc}
\includegraphics[width=0.425\linewidth]{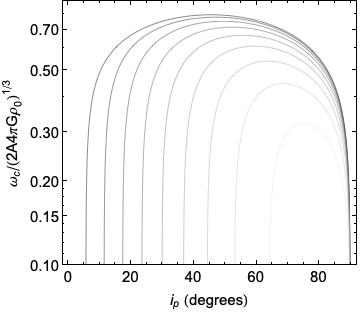}&
\includegraphics[width=0.425\linewidth]{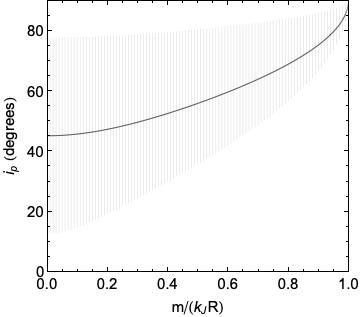}
\end{tabular}
\end{center}
\caption{(Left) The normalized growth rates of secondary spiral instabilities above the conventional $Q_M=1$ threshold at four different values of $m/(k_JR)$, from  $m/(k_JR)$=0.2 (black) to 0.8 (light gray) in steps of 0.2. 
(Right) The critical pitch angle $i_p=90\degree-\gamma$ where growth is maximized as a function of $m/(k_JR)$ (black line).  The width of the growth plateau, where the growth is above 75\% of the its peak value, is shown as a gray band.
}
\label{fig:growthShear}
\end{figure*}
In this section we return to the galaxy (non-rotating) frame and seek the spiral features where this process is most effective.  We focus on the earliest stages of growth, for which $k_e\rightarrow k$. The behavior and the spiral characteristics that we highlight are derived in the 3D context, but these are almost indistinguishable when treated in a 2D scenario, as discussed at the end of this section ($\S$~\ref{sec:2dv3d}).  
In addition, we discuss the implications of our calculations in the context of both gas disks and self-gravitating stellar disks, interchanging $1/k_J$ with the vertical scale height $h=\sigma/(4\pi G\rho_{0})^{1/2}$ in the latter case to obtain an approximate view of the scales and orientations where stellar spirals grow.  It should be noted, however, that phase mixing in stellar disks rather than pressure acts to counter self-gravity on small scales \citep[e.g.][]{kalnajs,JT}.  
 \subsection{Growth exclusive to trailing spirals}
One of the defining characteristics of growing open spirals that can be deduced from eq. (\ref{eq:omegacrossshear}) is that they are exclusively trailing above the Jeans length.  
This is intimately related to the role of the donkey effect (or inverse Landau damping) in the wave amplification process, which was shown in $\S$~\ref{sec:donkey} to lead to wave growth preferentially when the spiral is trailing.  The same is true for steady spirals, in order to communicate energy and angular momentum outward through resonant exchanges \citepalias{LBK}.   

Observed spirals are also characteristically trailing in galaxies.  For this reason we focus the predictions of our framework exclusively to this trailing regime, with the goal of enabling future comparisons with observations and numerical simulations.  In doing so we neglect predictions for the leading waves that eq. (\ref{eq:omegacrossshear}) predicts are capable of growth in gas disks exclusively below the Jeans scale.  This behavior is absent for stellar disks since it is purely a consequence of the pressure in the gas.  As $\vert k\vert\rightarrow \infty$ in stellar disks, the weakening of self-gravity due to phase mixing saturates so that $s_0^2\rightarrow 0$, rather than growing proportional to $k^2$ \citep[see e.g.][Figure 6.14]{BT08}.  Considering that this also effectively suppresses large-$k$ structures in combined gas and stellar disks \citep{rafikov}, we speculate that in real galaxies sub-Jeans leading gas structures never grow to prominence.  

\subsection{The growth rates and orientations of open spirals}
\label{sec:growthrates}
For the trailing spirals that grow through the donkey effect, the sensitivity to spiral forcing places constraints on the orientations where growth is fastest.  
Spirals with large $m$ and $k$ maximize their gravitational forcing, but gas pressure imposes a strict limit on how large $k$ and $m$ can be.  The result is a trade-off that brings the separations between spiral arms as close as possible to the Jeans length $\lambda_J=2\pi/k_J$ without dropping below it.  As the most straightforward of examples, spirals in the regime $Q_M\gtrsim 1$ achieve neutral stability when $k^2+m^2/R^2=k_J^2$ (or the more trivial scenario that either $m$=0 or $k$=0) (see eq. [\ref{eq:omegacross}] or e.g. eq [\ref{eq:omegacrossshear}]).  

To assess the conditions for genuine growth ($\omega_{i}>0$ or $-\omega_{c}^{3}>0$) it is convenient to consider shearing patterns and differentially non-steady modes separately.  In the first case, $m$ is presumed to be fixed while $k$ and the spiral orientation $\gamma$ evolve with time as the pattern shears.  
Spirals that are modes, on the other hand, are described with fixed $k$ and $m$.  
  
 \subsubsection{Amplifying shearing spirals}
For shearing spirals, we reexpress eq.~(\ref{eq:omegacrossshear}) in terms of $m/R$ and the spiral orientation $\gamma$, defined by $\tan\gamma=kR/m$, i.e.  
\begin{eqnarray}
&&Im(\omega)=\\
&=&\left[4\pi G\rho_02A\tan \gamma\left(\frac{1}{1+\tan^2\gamma}-\frac{m^2}{k_J^2R^2}\right)\right]^{1/3}\nonumber
\end{eqnarray}
The left of Figure \ref{fig:growthShear} shows the growth rate as a function of $i_p=90\degree-\gamma$ for different values of $m/(k_JR)$.  
The growth rate is also shown as a function of $i_p$ at different $k/k_J$ in the right panel for reference, but this will be discussed only later.  The behavior in both plots toward $k\rightarrow 0$ should be interpreted with caution, given that the calculations are meant to apply in the regime $kR\gg1$. 

The trends in $Im(\omega)$ offer a clear depiction of how the strengthening of gravitational force with increasing $k$ and decreasing $i_p$ brings about enhanced growth, until eventually pressure is able to dominate and suppress growth for the highest $k$ (lowest $i_p$).  The interplay of these two factors leads to a plateau of growth at intermediate pitch angles and maximum growth at a specific critical orientation.  We solve for the critical orientation by 
determining where $d\omega_{c}^3/d\tan\gamma=0$.  
This involves solving a quadratic equation for $\tan\gamma^2$ to find
\begin{equation}
\tan \gamma^2=-\left(\frac{k_J^2R^2}{2m^2}+1\right)\pm\frac{k_J^2R^2}{2m^2}\sqrt{1+8\frac{m^2}{k_J^2R^2}}
\end{equation} 
Only the upper of these yields real solutions.  

The right panel of Figure \ref{fig:growthShear} shows the critical orientation and marks the width of the growth plateau, taken from where the growth rate is at 75\% of its maximum value.  For these shearing spirals, the plateau width sets the length of time the spiral remains prominent as it swings in orientation.  The figure therefore suggests that spirals with lower arm multiplicity remain prominent for longer. 
A clear implication is that low-$m$ spirals are more likely to be observed over a range of orientations compared to spirals with a higher number of arms.  The latter class of objects would have comparatively higher uniformity, given the narrow spread in orientation at high $m$.  

\subsubsection{Amplifying spiral modes}
\begin{figure*}[t]
\begin{center}
\begin{tabular}{cc}
\includegraphics[width=0.425\linewidth]{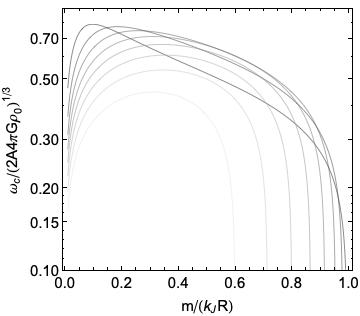}&
\includegraphics[width=0.425\linewidth]{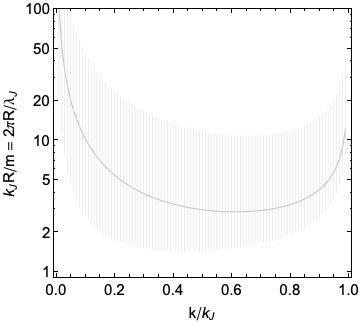}
\end{tabular}
\end{center}
\caption{(Left) Normalized growth rates as a function of $m/(k_JR)$ for non-axisymmetric `open' spiral instabilities above the conventional $Q_M=1$ threshold at different $k/k_{J}$ (black to light gray: from $k/k_{J}$=0.1 to 1 in steps of 0.1).  
(Right) The number of Jeans lengths that fastest-growing spiral arms are spaced azimuthally (black; $k_JR/m$) as a function of $k/k_J=\lambda_J/\lambda_R$.  
Spiral across a wide range of $k/k_J$ are spaced at roughly 3 times the Jeans length.
}
\label{fig:growthkm}
\end{figure*}
The growth and prominence of spirals is also meaningfully assessed in terms of $k$ and $m$, especially considering the possibility that perturbations triggered by processes in the galaxy will have either specific sets of $k$ and $m$, or be seeded as a spectrum of modes at a specific $k$.  This description can also be useful for shearing spirals, when the perturbing sources are very short lived such that the perturbation is removed before it has completed a broad swing.  

With this in mind, we express the growth rate as
\begin{equation}
Im(\omega)=\left[4\pi G\rho_02A\frac{m}{k_JR}\frac{k}{k_J}\left(\frac{1}{\left(\frac{k}{k_J}\right)^2+\left(\frac{m}{k_JR}\right)^2}-1\right)\right]^{1/3}
\end{equation}
(i.e. eq. [\ref{eq:omegacrossshear}] at times when $k_e\sim k$).
As shown in the left panel Figure \ref{fig:growthkm}, a plateau of growth is present over a range in $m/(k_JR)$ but the plateau narrows as $k/k_J$ decreases and the peak growth shifts to low $m$.  The behavior in the peak is expressed by
\begin{equation}
\frac{m_{crit}^2}{k_J^2R^2}=-\left(\frac{1}{2}+\frac{k^2}{k_J^2}\right)\pm\frac{1}{2}\sqrt{1+8\frac{k^2}{k_J^2}}  \label{eq:mcrit}
\end{equation} 
derived by solving for where $d\omega/\partial m$=0. Only the upper of these yields real solutions, which are allowed only as long as $k<k_J$.  

The right of Figure \ref{fig:growthkm} shows the inverse of the critical $m_{\rm crit}/(k_JR)$ given by eq. (\ref{eq:mcrit}), or $X_J=(k_JR)/m_{\rm crit}=(2\pi R/m_{\rm crit})/\lambda_J$, which measures the number of Jeans lengths that the spiral arms are spaced azimuthally.  The value $X_J$ is analogous to the swing parameter $X=k_{crit}R/m$ where $k_{crit}=\kappa^2/(2\pi G\Sigma)$ \citep[i.e.][]{toomre81, dobbsbaba, sellwoodmasters}.  The figure also shows the width of the plateau in $m_{\rm crit}/(k_JR)$, estimated based on where the growth rate is at 75\% of its peak value.  

With the exception of the sharp increase for $k/k_J\lesssim0.2$, the critical spacing is remarkably constant across over a range in $k/k_J$, with a value near 3.  Using eq. (\ref{eq:mcrit}) we locate the minimum value to $X_J\sim2.83$ at $k/kJ=\sqrt{3/2}/2$.  This can reach as low as 1.5-2 allowing for slightly slower growth.    

Given that $X_J=X/Q_M$ or approximately $4X/Q_T^2$ (using that $Q_T^2\approx 4Q_M$; \citealt{meidt22}) a critical value $\sim 3$ corresponds to $X=0.8-1.7$ for $Q_T=1-1.5$.  Our framework thus predicts amplification under similar conditions as previous analytical and numerical calculations \citep{toomre81,athanassoula84,dobbsbaba}.

A number of other interesting conclusions can be drawn from Figure \ref{fig:growthkm}.  We will focus mostly on the trend above $k/k_J\sim0.2$, given that we expect our predicted growth rates (calculated in the limit $kR\gg1$) to become inaccurate as $k\rightarrow0$. 

First, the constancy of $(k_JR)/m$ suggest that there is not much dynamic range in the number of spiral arms expected in galaxies. In Milky Way mass stellar disks, for example, letting $h=1/k_J\sim$0.45 kpc, stellar spiral patterns will have 2-5 arms.  In colder disk components the typical arm number increases.  Taking $\lambda_J$=300 pc as typical of the inner molecular disks in galaxies \citep{meidt23}, dense gas spirals within 1-2 disk scale lengths will range in multiplicity from $m$=25-50. 
This resembles the spur features detected in nearby spiral galaxies \citep{lavigne,meidt23, williams23}, and generally indicates that gas disks are structured on much smaller scales than their stellar counterparts.   

This uniformity in arm number from galaxy to galaxy aside, Figure \ref{fig:growthkm} also depicts variation in $m_{crit}$ moving to the largest $k/k_J$, leading to possibly observable trends between $k$ and $m$ in relation to the properties of the disk.  

We first consider fixed $k$.  In this case, disks with smaller $\lambda_J$ (larger $k_J$) become most unstable to perturbations on angular scales closer to the Jeans length than in warmer disks.  This would be visible as a trend in azimuthal arm spacings with global gas properties, and might also lead to systematic variations within disks (as long as $k$ varies considerably less across disks).  Thus, for example, arm number would decrease moving from the inner cold molecular disk to the outer HI reservoir in galaxies.  Systematic differences in the structures hosted by gas and stellar disks are also expected, even when both disks in a given galaxy are exposed to the same spectrum of perturbations.  According to Figure \ref{fig:growthkm}, these two disk components would favor growth in different parts of the spectrum; 
gas disks are more likely to host high-$m$ spur-like features, whereas stellar disks favor structures on larger angular scales.  

Interpreted in terms of variations at fixed $\lambda_J$, the right panel of Figure \ref{fig:growthkm} serves as another clear illustration of the trade-off to bring the separations between spiral arms as close to the Jeans length as possible: perturbations with smaller $\lambda_R=2\pi/k$ (toward the right) become preferentially unstable at larger typical angular separations (smaller $m$).  In other words, if $\lambda_R$ is already close to $\lambda_J$, the angular separation must be large, but an increase in $\lambda_R$ would give room for the angular separation to shift closer to the Jeans length. 

\subsection{The maximum unstable radial wavelength and a characteristic orientation}
\begin{figure}[t]
\begin{center}
\begin{tabular}{c}
\includegraphics[width=0.915\linewidth]{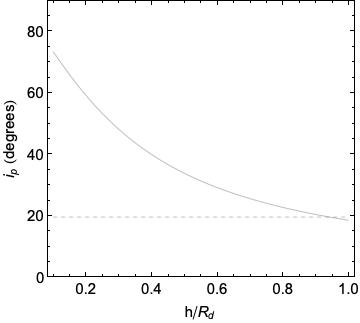}
\end{tabular}
\end{center}
\caption{Approximations for the maximum (solid line) and minimum pitch angles (dashed line) possible for growing spirals in disks with a range of $h/R_d$. These were estimated for $X_J=k_JR/m\approx3$ (see Figure \ref{fig:growthkm}).  }\label{fig:maxrange}
\end{figure}
As depicted in Figures \ref{fig:growthShear} and \ref{fig:growthkm}, through the Jeans length, the properties of disks regulate the spiral orientation or multiplicty of the dominant spiral perturbations at a given wavelength.  From this perspective, a key remaining ingredient for building predictions for how whole disks should appear given their properties is the most likely radial wavelength(s) present in those disks.  This could conceivably reflect the processes responsible for stimulating the perturbations (e.g. if factors in the disk seed perturbations with smaller $\lambda_R$ than larger-scale ex situ processes active within the gaseous, stellar or dark matter halo), and there may be reason to expect that the dominant sources vary systematically with global galaxy properties.  

On the other hand, we also expect the galaxy itself to act as an effective filter on the large number of the perturbations likely experienced by disks.  As discussed in the Appendix ($\S$~\ref{sec:longwave}), perturbations that extend beyond the disk scale length $R_d$ are stable as a rule.  
This defines the instability condition $\lambda_R<R_d$ that translates into a maximum possible orientation for prominent spirals in a given disk.  We illustrate this in Figure \ref{fig:maxrange}, where we have used the value $X_J=k_JR/m\sim 3$ (see Figure \ref{fig:growthkm}) as broadly representative of growing structures. An estimate for the minimum pitch angle $i_{p,min}$ is also shown there, determined from the same fixed $X_J$ using that $i_{p,min}=\arctan(X_J\sqrt{1-1/X_J})$. 

As disks become puffier and $h/R_d$ increases, the range of possible pitch angles is reduced given that
there is left less room below $R_d$ for disks to become unstable.  This quite naturally predicts that warmer disks will have a less complex appearance than colder disks, simply because there are fewer growing waves.  This could contribute to explaining why stellar disks tend to sustain only one dominant spiral pattern, whereas the gas disks in the same galaxy characteristically exhibit rich multi-scale structure.  By the same token, spiral structures can be completely suppressed in stellar disks that evolve to $h/R_d\sim 1$.  

\subsection{2D vs. 3D spirals}\label{sec:2dv3d}
\begin{figure*}[t]
\begin{center}
\begin{tabular}{cc}
\includegraphics[width=0.425\linewidth]{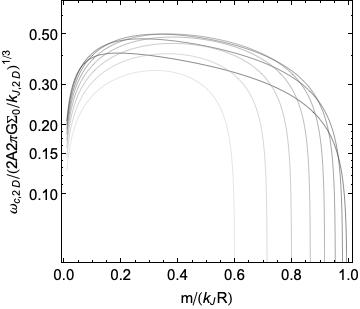}&
\includegraphics[width=0.425\linewidth]{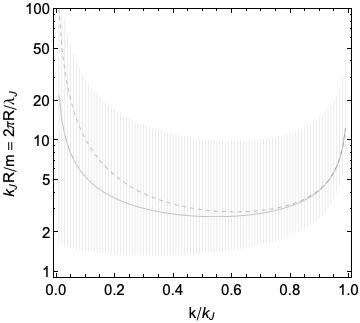}
\end{tabular}
\end{center}
\caption{(Left) Normalized 2D growth rates as a function of $m/(Rk_{J,2D})$ at different $k/k_{J,2D}$ (black to light gray: from $k/k_{J,2D}$=0.1 to 1 in steps of 0.1). (Right) Variation in the inverse of the 2D critical $m/(Rk_{J,2D})$ as a function of $k/k_{J,2D}$ (black line), together with width of the growth plateau (gray band). The 3D critical $m/(Rk_{J})$ is shown as a dashed line for reference adopting $k_{J,2D}=k_{J}$ (assuming that the disk is fully self-gravitating).  
}
\label{fig:critNJeans2D}
\end{figure*}
So far we have treated spirals as 3D perturbations that grow at the galactic mid-plane, but a related perspective envisions spirals as 2D objects.  Both gaseous and stellar disks are three-dimensional structures that can be expected to physically form either 3D or 2D spiral instabilities depending on whether or not they are perturbed in the vertical direction.  

Perhaps counterintuitively, within galaxies, the thinnest disk components are the most likely to experience 3D perturbations, since many of the processes taking place within the galaxy occur above a scale height.   Star-forming gas disks embedded in stellar disks and extended gas reservoirs, for instance, may be more likely to form 3D instabilities than stellar disks. 

Whether spirals are primarily 2D or 3D might also conceivably depend on whether in situ or ex situ processes are ultimately responsible for their origin.  For this reason it becomes relevant to seek measurable differences between the two.  
In the conventional (in)stability regime, \cite{meidt22} suggested 3D and 2D instabilities can be observationally distinguished  given that the former exist even after the threshold $Q_T=1$ is passed (until $Q_T\sim 2$).  One of the defining characteristics of structure formation outside the tight-winding limit, though, is that conventional thresholds no longer apply: non-axisymmetry is pervasive, with very little restriction in either 2D or 3D except a firm requirement on the orientation at a given $m$ or $k$ that replaces a traditional $Q$ threshold.  

In practice, differences in structural properties must serve as the basis for distinguishing between 2D and 3D spirals. These differences can be found by seeking the 2D effective dispersion relation, which follows from integration of the 3D dispersion relation over the vertical direction. In place of the 3D perturbation in $\S$~\ref{sec:framework} (eq. [\ref{eq:pertex}]), however, for 2D calculations it is  standard to adopt a vertical delta function perturbation $\rho_1=\Sigma_1\delta(z)$ in which case it is typical to let 
\begin{equation}
\Phi_1=\frac{2\pi G\Sigma_1}{\vert k_t\vert} e^{-k_tz}
\end{equation}
using Poisson's equation (see Binney \& Tremaine 2008) with
\begin{equation}
k_t=\sqrt{k^2+\frac{m^2}{R^2}}
\end{equation}
and neglecting disk thickness \citep[but see, e.g.][]{romeo92,rafikov, ghoshjog1}.

In this case, the 2D non-axisymmetric effective dispersion relation is 
\begin{eqnarray}
(\omega_e-m\Omega)^3&=&(\omega_e-m\Omega)\omega_{0,2D}^2+i\omega_{c,2D}^3\label{eq:disprelation2D}
\end{eqnarray}
where
\begin{equation}
\omega_{0,2D}^2=\kappa^2+\left(k_e^2+\frac{m^2}{R^2}\right)s_{0,2D}^2
\end{equation}
and
\begin{equation}
\omega_{c,2D}^3=2A\frac{km}{R}s_{0,2D}^2\label{eq:omegacross2d}
\end{equation}
in terms of 
\begin{equation}
s_{0,2D}^2=-2\pi G\Sigma_0\left(\frac{1}{\sqrt{k_e^2+\frac{m^2}{R^2}}}-\frac{1}{k_{J,2D}}\right).
\end{equation}
Here 
\begin{equation}
k_{J,2D}=2\pi G\Sigma_0/\sigma^2=hk_J^2
\end{equation}
writing $\Sigma_0=\rho_0(2h)$ \citep[e.g.][]{koyama,meidt22}. 

It is straightforward to show that solutions for $(\omega_e-m\Omega_p)$ are dominated by $\omega_{0,2D}$ below $Q_T=1$ and by $\omega_{c,2D}$ above $Q_T=1$.  
The left panel of Figure \ref{fig:critNJeans2D} depicts the growth rates implied in the latter scenario while the right panel shows the equivalent of Figure \ref{fig:growthkm}, now based on the 2D growth rates.  The critical 2D $X_{J,2D}$ behaves similarly to $X_J$ and is notably uniform over a large range in $k/k_{J,2D}$, with a value  $\sim 3$.  The structural properties and spacings of 2D and 3D spirals are therefore mostly similar.  However, measurable differences can emerge as disks become more weakly self-gravitating, given that $k_{J,2D}=k_J^2h$. 
The 2D and 3D Jeans lengths become increasingly different when the disk is embedded in a stronger background disk, in which case $h$ shifts below $\lambda_J$ ($hk_J<1$) and $\lambda_{J2D}>\lambda_J$ \citep{meidt22,meidt23}. 
 
The growth rates of 2D and 3D spirals are likewise mostly similar, though again with differences emerging as disks become less fully self-gravitating.  By comparing eq. (\ref{eq:omegacross2d}) with  $\omega_{c}^3$ in eq. (\ref{eq:omegacross}) we see that the growth rates of 2D spirals are $(h k_J)^2$ times the maximum 3D growth rates.  This leads to the slightly faster growth of 2D spirals in weakly self-gravitating disks (where $hk_J<1$).  Given that such fast-growing waves will also saturate faster (see $\S$~\ref{sec:decay}), however, 2D spirals do not necessarily dominate disk morphology more than 3D features.

\subsection{Discussion}
\label{sec:4discussion}
The predictions for the structural properties of spirals discussed in this section have several practical applications that make them relevant on a number of fronts.

First, the properties of spirals -- angular rotation speed, number of arms and orientation -- are important factors determining the impact these structures have on their host galaxies over short and long timescales.  This by now mature topic includes the outward transfer of energy angular momentum supported by resonant mode growth (\citetalias{LBK}, \citealt{GT,mark76,bertin78}), disk heating through transience and radial migration through resonant interactions \citep{fuchs3,sellwoodbinney,BT08}, non-circular streaming motions in gas and stars \citep[e.g.][]{robertsstewart,baba13,baba16, kawata, grand15motions}, azimuthal offsets between old and young stars and the natal gas \citep[i.e.][]{fujimoto,roberts69,dobbsspirals,pettitt17}, gas inflows driven by gravitational torques \citep{block02,hopkins11,bournaud11}, azimuthal variations in stellar and gas phase metallicity \citep{grand15,grand16,Khoperskov18}, etc.    The ability to predict which spirals are likely to be present at any given time makes this framework an important step toward improved consistent modelling of spiral-driven secular evolution in the context of semi-analytical or numerical models of galaxy evolution \citep[e.g.][]{krumBurk, forbes12,forbes14,cacciato1,genel12}

Second (and relatedly), by predicting a testable link between spiral properties and perturbers and disk conditions, our framework promotes a quantitative approach to characterizing spirals in observations \citep[e.g][]{elmegreen80, kennicutt81, seigar98, grosbol04, merrifield06,kendall15,hart17,hart18,hunt18,masters18, yu19,kwak19}, dedicated simulations \citep{li05,fujii11,grand12,minchev12,roskar12,donghia13,kumamoto16,michikoshi16,pettitt16,SC,michikoshi18,fujii18,derijcke} and cosmological numerical simulations \citep{goz15,mollitor15,snyder15,grand17,mutlu18,sanderson20}, thus potentially helping in the longstanding debate about their origin and nature (e.g. \citealt{linshu66,JT,laubertin} and reviews by \citealt{dobbsbaba} and \citealt{sellwoodmasters}).  
More tangibly in this context, our results strongly favor the transient spiral picture being developed in the literature \citep{sellwoodlifetimes,SC14,SC,SCnonlin,SCsat}; whether spirals are shearing material patterns or growing modes, their amplification acts as its own off switch (see $\S$~\ref{sec:decay}, $\S$~\ref{sec:donkey} and \citealt{hamilton}).  This makes it likely that any genuine steady spiral modes are also short-lived rather than the quasi-stationary features envisioned by \cite{linshu66}.  

Beyond the question of spiral lifetimes, our calculations suggests that it is difficult to discern whether spirals are modes or shearing in nature based on structural properties alone, as considered by \cite{yu19}, \cite{pringle19} and \cite{lingard21}, since both amplify most efficiently with the same properties.  Distinguishing one type from the other must take one of the many other techniques devised for this purpose \citep[see reviews by][]{dobbsbaba,sellwoodmasters}.  
When orientation is being examined, we find it noteworthy that, although the orientations of shearing spirals are locally determined by the shear rate, globally there is no predicted link between the most favorable orientation and galactic shear from galaxy to galaxy.  Our calculations instead suggest a link with local velocity dispersion and surface density (i.e. through the Jeans length or stellar scale height) and also the disk scale length.

In the context of gas disks specifically, the rich multi-scale structure predicted by our framework has a number of implications.   First, the capability of disks (both nearby and at earlier cosmic times at high redshift) to form structures through local gravitational instability, and ultimately stars, is commonly assessed using the Toomre $Q_T$ parameter \citep[e.g][]{martin01,boissier03,cacciato12,genzel13,maseda14,westfall14,ubler19}, which was derived for structures in the axisymmetric and tight-winding limits.  However, disks are susceptible to the growth of non-axisymmetric structures even near or above conventional thresholds as a result of swing amplification   (demonstrated within our new framework in \S\ref{sec:shortwaveinstability}; also see \citetalias{GLBb} and \citet{JT}),  making the applicability of $Q_T$ as a diagnostic of collapse and star formation unclear.  
Second, non-axisymmetric instabilities are predicted to contribute to ISM morphology over a range of scales, down to roughly the Jeans length.  
This makes it a relevant addition to 
the picture of multi-scale gas structure being actively developed and observationally tested today, which 
includes spur formation through instabilities that take place along the high density arms themselves \citep[e.g.][]{elm89,ko06,wadakoda,mandowara,sormani} and coordinated feedback \citep[e.g.][]{kko20}.  We suggest that the rapid structure formation predicted by our framework is an important ingredient for rebuilding the reservoir for continued star formation given the continual destruction of structure via feedback \citep[e.g.][]{elm11,maclow17,kim21fFB,chevance,kim22,watkins23} and also for driving turbulence over a range of scales \citep{meidt18,meidt20,meidt22}.  

\section{Summary and conclusions}
In our modern view of galaxy evolution, gas and stellar disks are embedded in highly structured stellar and dark matter halos in contact with evolving multi-phase gas reservoirs \citep[e.g.][]{moore99,ferguson02,sancisi08,bonaca19} and this keeps them nearly constantly exposed to a rich spectrum of perturbations.  From this perspective, explaining the multi-wavelength spiral morphologies of galaxies is arguably less a matter of how any one perturbation originates and more a question of  
 why disks favor the growth of a finite subset of these perturbations.  
 
 In this paper we propose an approach that makes it possible to predict the properties of the spirals that evolve to prominence under a given set of conditions.  Given that spirals are often observed with orientations that put them near or beyond the edge of the tight-winding limit (which requires $i_p\ll  45\degree$), we need to push beyond standard Lin-Shu theory.  This shifts us into the regime of swinging shearing patterns \citep{GLBb,JT,toomre81}.  But rather than keep all the functionality of density-wave theory separate from swing-amplification, as is typical in spiral structure theory, 
 in this paper we propose an approach that involves merging 
the Lin-Shu wave framework with the idea of shearing and non-steady spirals.  We then apply it (for simplicity) within the context of an idealized gas disk in hydrostatic equilibrium.  

This is the first time these two main branches of spiral structure theory have been joined together into a single framework, and we find that there are a number of advantages:
 \begin{enumerate}
\item The stability of any spiral can be described with this one framework (\S\ref{sec:2}), regardless of whether it is a density-wave or a swing-amplifying material pattern. This means that, to make predictions, it is not necessary to know a priori how the perturbation originated, whether in situ, during the passage of a companion galaxy, or as a mode propagating between resonances. 
 
\item The expression we derive (\S\ref{sec:3dispersionrelation} and Appendix) to describe the evolution of the perturbation  acts as an effective dispersion relation (though it is not a genuine one).  This makes it capable of capturing not only conventional instability in the tight-winding limit (\S\ref{sec:conventionalQ}), but also a transition -- once the conventional stability threshold $Q_M=1$ (or $Q_T\approx 1$) is passed -- to a regime in which disks remain ultra responsive to non-axisymmetric structures (\S\ref{sec:shortwaveinstability}).  This means that in one expression, we describe tight-winding density waves as well as swing-amplifying shearing spirals and the resonant and non-resonant growth of `open' spiral wave modes.   

\item In the Lin-Shu-centered framework, radial and azimuthal spiral forcing are conspicuous factors in the effective dispersion relation, even in the swing-amplification regime.  This makes it straightforward to predict the spiral orientations most favorable to amplification (\S\ref{sec:bottomsdream}) and also means that the swing-amplification process can be re-examined as a more general phenomenon.  

 \end{enumerate}

As an illustration of why spiral forcing is so important, we complement our derivation of the effective dispersion relation with a study of motion in the spiral arm frame (\S\ref{sec:spiralarmframe}), specifically looking separately at the radial and azimuthal directions.  This keeps the forcing terms relevant for amplification clear and showcases, in particular, the importance of the donkey effect (or inverse Landau damping) discussed by Lynden-Bell.  The latter of these becomes relevant especially in the regime of open spirals where it is important not only for mode growth but also for swing-amplification, which can be thought of as a cousin of non-resonant mode growth.  We propose referring to the amplification in this regime and its cessation, generally, as originating with the donkey effect (or `Bottom's Dream').  

Growth through the donkey effect entails the mechanism for its own suspension.  This can be the result of either nonlinear saturation \citep{hamilton} or a weakening of self-gravity as a consequence of the growth ($\S$~\ref{sec:decay}) present in the linear regime.  In either case, the growing wave is no longer able to impart the motion necessary to support itself.  
Thus, the amplifying spirals predicted in this framework are transient features, with lifetimes lasting several orbital periods at most.  

We use the effective dispersion relation to derive an analytical expression (\S\ref{sec:growthrates}) for the orientations of the fastest-growing (most prominent) spirals in terms of local conditions, adding a quantitative, testable prediction that has been missing from the literature.  
The competition between spiral forcing, which increases as spiral segments become more closely spaced, and gas pressure restricts amplification to a specific set of spiral orientations.  
These critical orientations are the same regardless of the nature of the spiral arms (all other factors held fixed).

We discuss the implications of these predictions in the context of spirals in gas disks and stellar disks, equating the Jeans length with the stellar scale height to deduce approximate behavior.
We suggest (\S\ref{sec:4discussion}) that: 1) stellar disks favor prominent two- and three-armed spirals over a broad range of $i_p$, down as far $i_p=10\degree$, similar to observations; 2) dynamically cool gas disks favor more complex structure with a range of $m$ present over a broader range of scales, including near the Jeans length.  The highest multiplicity spiral arms that constitute the small scale structure in gas disks would exhibit a considerably narrower range of pitch angles $i_p>45\degree$ than characteristic of lower $m$ structures, making them similar to gas spur features observed in emission and extinction in nearby galaxies.

The azimuthal spacings of many prominent structures are similar and characteristically only a few times the Jeans length. 
Consequently, the radial wavelength of a given mode could serve as a useful diagnostic of how it originated, e.g. if long (short) modes are characteristic of external (internal) perturbers. It might also provide an indirect constraint on the Jeans length or scale height.  

As a result of characteristic stability on scales larger than the disk scale length $R_d$, spirals with radial wavelengths longer than $R_d$ decay rather than grow.   This sets a maximum maximum pitch angle above which disks no longer support spiral patterns.  
We suggest that this is a natural way to describe the observed stellar spiral pitch angle distribution, which decreases and narrows with increasing stellar mass (or correlated global galaxy property).  It is also a first-principles explanation for the suppression of shearing spiral structure in the puffiest (dynamically hottest) disks; as the gap between the Jeans length and $R_d$ decreases and even disappears, there are no viable scales left for the disk to become gravitationally unstable (\S\ref{sec:4discussion}).     

Our new framework motivates future observational and numerical studies of spiral patterns.  Perhaps most immediately, we offer an interpretation of the rich structure observed with JWST/MIRI on small scales in the dusty disks of nearby galaxies:
gravitational instability is commonplace on scales extending well below the typical scales of grand-design spiral patterns, making it a mechanism capable of forming highly regular and elongated gas spurs observed in bars and spirals.  This also makes gravitational instability a viable source of turbulence observed on small scales in gas disks, with fast enough growth rates to replenish the dense gas structures that are just as quickly destroyed by star formation feedback. From this perspective, gravitational instability and swing amplification in particular play a key role in organizing the cold ISM and the process of star formation.\\

\noindent Acknowledgements\\
We would like to thank the referee for their careful review and helpful comments that helped improve the clarity of the paper.  S. Meidt would also like to thank S. de Rijcke for a discussion of some of the recent literature on spiral structure. 

\appendix
\restartappendixnumbering
\section{The 3D effective dispersion relation: derivation and characteristics}
\subsection{Derivation}
\label{sec:appendix_derivation}
To derive an expression that behaves like a dispersion relation for generic non-axisymmetric wave perturbations capable of growth, we begin by substituting the potential perturbations in eq. (\ref{eq:pertex}) and eq. (\ref{eq:perturbations}) introduced  in $\S$~\ref{sec:framework} into the linearized equations of motions 
\begin{equation}
i(-\omega_e+m\Omega)v_{R,1}-2\Omega v_{\phi,1}=-\frac{\partial\Phi_1}{\partial r}-\frac{\sigma^2}{\rho_0}\left(\frac{\partial\rho_1}{\partial r}\right)\label{eq:radialforce}
\end{equation}
\begin{equation}
i(-\omega_e+m\Omega)v_{\phi,1}-2B v_{R,1}=-\frac{1}{R}\frac{\partial\Phi_1}{\partial \phi}-\frac{\sigma^2}{\rho_0}\left(\frac{1}{R}\frac{\partial\rho_1}{\partial \phi}\right)\label{eq:azimuthalforce}
\end{equation}
\begin{equation}
i(-\omega_e+m\Omega)v_{z,1}=-\frac{\partial\Phi_1}{\partial z}-\frac{\sigma^2}{\rho_0}\left(\frac{\partial\rho_1}{\partial z}\right).
\end{equation}
We then solve for  $v_{R,a}$, $v_{\phi,a}$ and $v_{z,a}$ finding
\begin{eqnarray}
v_{R,a}&=&-\frac{\chi_a}{\Delta}
\left(k_e(\omega_e-m\Omega)+i\frac{2m\Omega}{R}\right)
\label{eq:radvelocity}
\end{eqnarray}
and
\begin{eqnarray}
v_{\phi,a}&=&-\frac{\chi_a}{\Delta}\left(2Bik_e+\frac{m(\omega_e-m\Omega)}{R}\right)
\label{eq:phivelocity}
\end{eqnarray}
in the limit of negligible vertical rotational lag (Meidt 2022), and
\begin{equation}
v_{z,a}=-\frac{k_z\chi_a}{(m\Omega-\omega)}+i\frac{(\nabla \mathcal{F}+\sigma^2\frac{\nabla \mathcal{R}}{\rho_0})}{(m\Omega-\omega)}e^{ikR+ik_z z}.
\label{eq:vertvelocity}
\end{equation}
Here 
\begin{eqnarray}
\Omega&=&V_c/R\nonumber\\ 
B&=&-\Omega-\frac{1}{2}R\frac{d\Omega}{dR}\\\nonumber
\Delta&=&\kappa^2-(m\Omega-\omega_e)^2\\\nonumber
\kappa^2&=&-4B\Omega.\nonumber
\end{eqnarray}
and
\begin{equation}
k_e=k-\int\frac{\partial\omega}{\partial R}dt,
\end{equation}
\begin{equation}
\omega_e=\omega-R\frac{\partial k}{\partial t},
\end{equation}
\begin{equation}
\chi_a=\left(\Phi_a+\sigma^2\frac{\rho_a}{\rho_0}\right).
\end{equation}
in terms of the isotropic isothermal velocity dispersion $\sigma$ ($\S$~\ref{sec:framework}).  

We work in a regime in which quantities in the equilibrium disk vary with radius but only slowly evolve in time, such that the time derivatives of those quantities are neglected.  The perturbation's pattern speed (the real part of $\omega$) is also assumed to vary much more slowly in time than the wave's other properties.  

According to Poisson's equation (see eq. \ref{eq:poisson}), 
\begin{equation}
\Phi_a=\frac{4\pi G\rho_a}{(ik_e+T_R)^2+\frac{(ik_e+T_R)}{R}-\frac{m^2}{R^2}+(ik_z+T_z)^2}\label{eq:poissonFull}
\end{equation}
where $T_R$ and $T_z$ represent the characteristic scales of the perturbation's amplitude variations in the radial and vertical directions, $T_R=dln\mathcal{F}(R,z)/dr$ and $T_z=dln\mathcal{F}(R,z)/dz$.  

These expressions for the perturbed velocities are ultimately substituted into the perturbed continuity equation, which  
reads
\begin{equation}
\frac{\partial\rho_1}{\partial t}+\frac{1}{R}\frac{\partial R\rho_0 v_R}{\partial R}+\frac{1}{R}\frac{\partial\rho_0 v_\phi}{\partial \phi}+\Omega\frac{\partial\rho_1 }{\partial \phi}+\frac{\partial\rho_0 v_z}{\partial z}=0.\label{ref:linearizedCont}
\end{equation}
The second (radial) term on the left hand side has three terms
\begin{equation}
\frac{1}{R}\frac{\partial R\rho_0 v_R}{\partial R}=\rho_0 \frac{v_{R,1}}{R}+\rho_0 \frac{\partial v_{R,1}}{\partial R}-\rho_0 \frac{v_{R,1}}{R_d}
\end{equation}
where $R_d=-(\partial\ln\rho_0/\partial R)^{-1}$.  
Assuming that the radial variation of the perturbation is much faster than quantities in the equilibrium disk ($kR\gg1$ and $kR_d\gg1$), the first and third terms on the right are negligible and the second term simplifies to $ik\rho_0 v_{R,1}$ (see also \citealt{GLBb}) since
\begin{equation}
\frac{\partial v_{R,1}}{\partial R}=\frac{\chi_a}{\Delta}\left(-ik_e^2(\omega_e-m\Omega)+\frac{k_em2\Omega}{R}\right)
\end{equation}
in this limit.  The continuity equation in this case, after substitution of the expressions for the perturbed radial, azimuthal and vertical velocities (eqs. \ref{eq:radvelocity}, \ref{eq:phivelocity} and \ref{eq:vertvelocity}), becomes
\begin{eqnarray}
0=-i(\omega_e-m\Omega)\rho_1+\frac{\partial\rho_0 v_z}{\partial z}&+&\rho_0\frac{\chi_1}{\Delta}\left[-i\left(k_e^2+\frac{m^2}{R^2}\right)(\omega_e-m\Omega)+\frac{2Ak_em}{R}\right] \label{eq:contFast1}
\end{eqnarray}
using that $2A=2\Omega+2B$.  Here we have also assumed that $\omega_e\gg\dot{k}_e/k_e$, to focus on instantaneous or fast growth.      

We rewrite eq. (\ref{eq:contFast1}) as
 \begin{eqnarray}
0=& -&(\omega_e-m\Omega)^3\nonumber\\
 &+&\Bigg[\kappa^2+\left(k_e^2+\frac{m^2}{R^2}\right)s_0^2+C_z \Delta\Bigg](\omega_e-m\Omega_p)\nonumber\\
&+&i s_0^2\left(\frac{2Ak_em}{ R}\right),\label{eq:full3drelationshort}
\end{eqnarray}
obtaining the dispersion relation in the `short-wave' limit discussed in the main text, where
\begin{equation}
s_0^2=\chi_a\frac{\rho_0}{\rho_a},
\end{equation}
\begin{equation}
\Phi_a=\frac{-4\pi G\rho_a}{k^2+\frac{m^2}{R^2}-(ik_z+T_z)^2}
\end{equation}
(eq. \ref{eq:poissonFull} in the WKB approximation) and  
\begin{equation}
C_z=i\frac{\partial \rho_0 v_z}{\partial z}\frac{1}{\rho_1(\omega_e-m\Omega)^2}.
\end{equation}
At the mid-plane of a disk in hydrostatic equilibrium, $T_z\rightarrow 0$ and the vertical term simplifies to
\begin{equation}
C_z\approx-\frac{1}{(\omega_e-m\Omega)^2} k_z^2s_0^2
\end{equation}
following the arguments in \cite{meidt22}.  This term makes the dispersion relation quartic in $(\omega-m\Omega)$.  \cite{meidt22} discusses how to solve it to obtain the conditions for stability 
under the assumption that $\partial\omega/\partial R=0$, i.e. neglecting the last term on the right hand side in eq. (\ref{eq:full3drelationshort}). In practice, however, it is more straightforward to argue that the fastest growing perturbations are those that satisfy $C_z\rightarrow 0$, i.e. by entailing negligible vertical phase variation below the scale height such that $k_z\ll k^2+m^2/R^2\ll 1/h^2$, as discussed in the main text. 

\subsection{The significance of the shear term for material patterns and amplifying modes}
\label{sec:appendix_shear}
To gain insight into the significance of the $2Akm/R$ term in eq. (\ref{eq:full3drelationshort}) it is worthwhile to consider it within the context of the
most general form of the continuity equation, which involves the radial velocity derivative
\begin{eqnarray}
\frac{\partial v_R}{\partial R}&=&\frac{\chi_1}{\Delta}(\omega_e-m\Omega)\left(-ik_e^2-\frac{k_e}{R}\left[\frac{\partial \ln(\omega_e-m\Omega)}{\partial \ln R}-\frac{\partial \ln\Delta}{\partial \ln R}\right]\right)\nonumber\\
&+&\frac{\chi_1}{\Delta}\frac{m2\Omega}{R}\left(k_e+i\frac{1}{R}\left[1-\frac{\partial \ln \Omega}{\partial \ln R}+\frac{\partial \ln\Delta}{\partial \ln R}\right]\right).
\end{eqnarray}
The term $\partial(\omega_e-m\Omega)/\partial R$ here encodes a link between wave growth and the factor $2Am$ that is the same for amplifying shearing patterns and for amplifying modes.  
Explicitly writing $\omega=m\Omega_p+i\omega_i$ as the sum of real and imaginary components, 
\begin{eqnarray}
\frac{\partial(\omega_e-m\Omega)}{\partial R}&=&\left(\frac{\partial\omega}{\partial R}-\frac{\partial(R\dot{k})}{\partial R}-m\frac{\partial\Omega}{\partial R}\right)\nonumber\\
&=&\left(\left[m\frac{\partial\Omega_p}{\partial R}-\frac{\partial(R\dot{k})}{\partial R}\right]+\frac{2Am}{R}+i\frac{\partial\omega_i}{\partial R}\right)\label{eq:partialOmega}.
\end{eqnarray}
It is significant that the first two factors in brackets on the right hand side are either exactly equal to zero or sum (nearly) to zero for modes and material patterns, respectively.  
In the latter case, the sum in the square brackets generally equals the small quantity $mRd\dot{k}/dR=-mR d^2\Omega_p/dR^2 $ since $\dot{k}=-md\Omega_p/dR$.  In the particular case that $\Omega_p=\Omega$, the term in square brackets equals
\begin{equation}
m\frac{\partial\Omega_p}{\partial R}-\frac{\partial 2Am}{\partial R}\approx \frac{2Am}{R}-(1-\beta)\frac{2Am}{R}=\beta\frac{2Am}{R}\nonumber,
\end{equation}
which is exactly zero where the rotation curve is flat ($\beta=0$).
This makes modes and shearing patterns ideal situations for growth, as is clear rewriting eq. (\ref{eq:partialOmega}) as 
\begin{eqnarray}
\frac{2Am}{R}&=&-i\frac{\partial\omega_i}{\partial R}+\frac{\partial(\omega_e-m\Omega)}{\partial R}-\left[m\frac{\partial\Omega_p}{\partial R}-\frac{\partial(R\dot{k})}{\partial R}\right]\label{eq:2am}\\
&\approx&-i\frac{\partial\omega_i}{\partial R}\label{eq:2amapprox}
\end{eqnarray}
where the latter equivalency applies especially in the `short-wave' limit where $\partial(\omega_e-m\Omega)/\partial R$ is negligible, whenever the term in square brackets approaches zero. 

The equivalence in eq. (\ref{eq:2amapprox}) characteristic of the short-wave limit implies both that 
\begin{equation}
k_e=\begin{cases}
k&\text{material spirals}\\
k+\int\frac{2Am}{R}dt&\text{spiral modes}
\end{cases}
\end{equation}
(as used in $\S$~\ref{sec:decay}) and that 
\begin{eqnarray}
\dot{k}&=&-m\frac{\partial\Omega_p}{\partial R}=\frac{2Am}{R}-m\frac{\partial(\Omega_p-\Omega)}{\partial R}\\
&=&-i\frac{\partial\omega_i}{\partial R}+m\frac{\partial(\Omega-\Omega_p)}{\partial R}, 
\end{eqnarray}  
from which we see that 
\begin{equation}
(\omega_e-m\Omega)=m(\Omega_p-\Omega)-mR\frac{\partial(\Omega_p-\Omega)}{\partial R}+i\left(\omega_i-R\frac{\partial\omega_i}{\partial R}\right). 
\end{equation} 

\subsection{The general dispersion relation}\label{sec:genDR}
Outside the short-wave limit, when the equivalence in eq. (\ref{eq:2amapprox}) no longer applies, the evolution of spiral perturbations is no longer tied solely to $\partial\omega_i/\partial R$.   For the general scenario in which the gradients in the disk are not necessarily negligible, we use eq. (\ref{eq:2am}) to write the continuity equation in its most general form as
\begin{eqnarray}
0=-i(\omega_e-m\Omega)\rho_1+\frac{\partial\rho_0 v_z}{\partial z}&+&\rho_0\frac{\chi_1}{\Delta}(\omega_e-m\Omega)\left\{-ik_e^2-i\frac{m^2}{R^2}+\frac{k_e}{R}\frac{\partial \ln\Delta}{\partial \ln R}-k_e\left(\frac{1}{R}-\frac{1}{R_d}\right)\right\}\nonumber\\
&+&\rho_0\frac{\chi_1}{\Delta}\frac{2m\Omega}{R}\left\{-i\left(\frac{1}{R}-\frac{1}{R_d}\right)+i\frac{1}{R}\left(1-\frac{\partial \ln \Omega}{\partial \ln R}+\frac{\partial \ln\Delta}{\partial \ln R}\right)\right\}\nonumber\\
&+&\rho_0\frac{\chi_1k_e}{\Delta}\left(-i\frac{\partial \omega_i}{\partial R}-\left[m\frac{\partial\Omega_p}{\partial R}-\frac{\partial(R\dot{k})}{\partial R}\right]\right), 
\label{eq:simp3Dv1}
\end{eqnarray}
which can be simplified further using that 
\begin{equation}
\frac{\partial \ln\Delta}{\partial \ln R}=2(\beta-1)+D\label{eq:delta}
\end{equation}
where
\begin{equation}
D=\frac{(\omega_e-m\Omega)}{\Delta}\left(2(\beta-1)\omega_e-2R\frac{\partial\omega_e}{\partial R}\right).
\end{equation}
Placing eq.(\ref{eq:delta}) in the context of eq. (\ref{eq:simp3Dv1}) makes clear that the waves that grow fastest are those that satisfy $D=0$, which is equivalent to assuming $\partial \ln{\omega_i}/\partial \ln R=(\beta-1)$. Fast growth might otherwise have negligible $\partial \ln\Delta/(\partial \ln R)$, for example. 

Under the assumption that $D\approx0$ and the term in square brackets in eq. (\ref{eq:simp3Dv1}) is negligible, we arrive at  \begin{eqnarray}
0=& -&(\omega_e-m\Omega)^3\nonumber\\
 &+&\Bigg[\kappa^2+\left(k_e^2+\frac{m^2}{R^2}-ik_e\left(\frac{3-2\beta}{R}-\frac{1}{R_d}\right)\right)s_0^2+C_z \Delta\Bigg](\omega_e-m\Omega)\nonumber\\
&+&i s_0^2\left(-ik_e\frac{\partial\omega_i}{\partial R}\right)+s_0^2\frac{m\Omega}{R}\left(\frac{1-\beta}{R}-\frac{1}{R_d}\right)\label{eq:full3drelation}
\end{eqnarray}
after substitution of expressions for the perturbed radial, azimuthal and vertical velocities (eqs. \ref{eq:radvelocity}, \ref{eq:phivelocity} and \ref{eq:vertvelocity}).  

Eq. (\ref{eq:full3drelation}) is equivalent to eq. (\ref{eq:full3drelationshort}) in the short-wave limit, where $2Am/R=-i\partial\omega_i/\partial R$ and factors of order $1/R$ and $1/R_d$ are negligible.  In the opposite `long-wave' limit, gradients in the disk are more important than the perturbation's phase variation, and the term proportional to $i\partial\omega_i/\partial R$ is outweighed by the second term in the bottom line of eq. (\ref{eq:full3drelation}).  

\subsection{The `long-wave' limit $kR_d\ll1$}\label{sec:longwave}
In the `long-wave' regime $k\rightarrow0$ and the dispersion relation in eq. (\ref{eq:full3drelation}) (assuming $D\approx 0$) reduces to 
\begin{equation}
(\omega_e-m\Omega)^3\approx\left(\kappa^2+\frac{m^2}{R^2}s_0^2\right)(\omega_e-m\Omega)+\frac{m}{R}s_0^2\left(\frac{1-\beta}{R}-\frac{1}{R_d}\right)\Omega\label{eq:longwave1}
\end{equation}
where now $s_0^2$ from eq. (\ref{eq:s0sq}) is written as
\begin{equation}
s_0^2\approx\frac{-4\pi G\rho_0}{\frac{m^2}{R^2}-T_R^2}+\sigma^2,  
\end{equation}
once again taken to apply at the galactic mid-plane favorable to instability.  

Specifically in the limit $Q_M>1$, eq. (\ref{eq:longwave1}) reduces to 
\begin{equation}
(\omega_e-m\Omega)^3\approx\frac{m}{R}s_0^2\left(\frac{1-\beta}{R}-\frac{1}{R_d}\right)\Omega, \label{eq:largeQlong}
\end{equation}
from which we infer a new set of conditions for growth.  Most relevant to this work, eq. (\ref{eq:largeQlong}) implies 
that waves with $T_R\sim1/R_d\ll m/R\ll k_J$ for which self-gravity dominates over pressure ($s_0^2<0$) are able to grow out to a maximum radius $R\sim R_d$ and, beyond this radius, growth is suppressed.  That is, for growth in the long wave regime we must have $1/R\gg k\gtrsim1/R_d$.  
We can thus conclude that open spirals grow only inside $\sim R_d$ and have wavelengths $\lambda_r=2\pi/k$ that are $\sim R_d$ at longest.  

\subsection{Comparison with other non-axisymmetric dispersion relations}\label{sec:otherRelations}
It is worth examining how the effective dispersion relation in eq. (\ref{eq:simp3Dv1}) relates to a number of other dispersion relations.
The 2D Lin-Shu dispersion relation, for one, is obtained from eq. (\ref{eq:simp3Dv1})  in the tight-winding limit $kR\gg m$ (also taking $kR_d\gg1$)  after integration over the vertical direction:  
\begin{equation}
(\omega-m\Omega)^2=\kappa^2-2\pi G\Sigma k+\sigma^2 k^2.
\end{equation}  
This is similar to the dispersion relation derived for stars by \cite{kalnajs} in the tight-winding approximation \citep[e.g.][eq. 6.61]{BT08}, without the third pressure term on the right.   

In the less strict limit $kR\gtrsim m$ of primary interest in this work, 
integration over the vertical direction yields a less spare version of eq. (\ref{eq:simp3Dv1}) that becomes equivalent to the expressions examined by \cite{laubertin, meidt22,GT}, depending on several choices.  

For example, following \cite{laubertin}, one would drop the imaginary terms appearing in eq. (\ref{eq:simp3Dv1}) or eq.  (\ref{eq:full3drelation}) based on their argument that they represent out-of-phase factors that are unimportant for local disk stability.\footnote{\cite{laubertin} also ignore a real factor proportional to $1/(\omega-m\Omega)$.  In our derivation, real and imaginary factors proportional to $1/(\omega-m\Omega)$ (i.e. after dividing eq. (\ref{eq:simp3Dv1}) by $(\omega-m\Omega)$) are the reason our dispersion relation is cubic instead of quadratic.  From the perspective of this work, these factors are key for instability beyond the conventional thresholds.  They appear in the third line of eq. (\ref{eq:full3drelation}).  One term is important in the short-wave limit, while the other regulates the features of $Q_M\gtrsim 1$ instability in the long-wave limit. }   

A slimmer version of the \cite{laubertin} relation that is quadratic in $(\omega-m\Omega)$ adopts the limit $kR_d\gg 1$ and/or $k\gg d \ln\Omega/dr$ \citep{morozov}, as implemented by \cite{meidt22}.  The latter choices are particularly relevant  
for open `short-wave' spirals with intermediate pitch angles.  
The limit $k\gg 1/R_d$ is consistent with the WKB approximation and, as in e.g. \cite{GT}, we invoke it together with the assumption that $T_R\rightarrow 0$.  
With this choice, a number of the terms identified by \cite{laubertin} (specifically their T1 or $\mathcal{J}$) are absent.\footnote{In their derivation of the local dispersion relation, \cite{laubertin} identify the $T_1$ factor appearing in their eq. (C14) as the source of enhanced instability.  This $T_1$ factor shares qualities in common with $\omega_{c}^3$ in eq. (\ref{eq:omegacross}) highlighted in section \ref{sec:relationtoswing}; it goes to zero both when $m=0$ and in the absence of rotation curve shear.  However, it appears that the combination of $T_1$ with the last two factors in their eq. (C14) together constitute the terms we identify with open spiral instability.  This is difficult to confirm given that the factor $A$ in their neglected imaginary term is never defined.  Their choice to expand out to second order in $m/(kR)$ (which they assume is small) may have motivated their separation of the $T_1$ term from the final term in  eq. (C14) proportional to $1/(\omega-m\Omega)$.}  At the same time, it is also convenient to assume $kR\gg(1-\beta)$, which even more strictly neglects variation in all quantities in the background equilibrium disk.  This can be invoked without losing the factors $m/R$ assuming, as in eq. (\ref{eq:full3drelationshort}) and in the main text, that $m\gg(1-\beta)$.  In practice, then, the form of the dispersion relation in the short-wave limit in eq. (\ref{eq:full3drelationshort}) applies in a slightly more local regime than the local dispersion relation derived by \cite{laubertin} and includes factors that they neglect (see footnotes 5 and 6).  

Once the dispersion relation is simplified to apply in the short-wave limit, as in eq. (\ref{eq:full3drelationshort}) or eq. (\ref{eq:full3drelation}), it can be reduced to a genuine non-axisymmetric dispersion relation, 
in the form derived by \cite{GT} specifically for the case that spirals are genuine wave modes of the disk.  (Lau \& Bertin derive their dispersion relation under a similar condition.)  Given our desire to remain as general as possible, however, we choose to 
keep all terms that 
involve variation in $\omega$ with radius ($d\omega/dR\neq0$) and explicitly include an imaginary growing component that accommodates an imaginary term in the dispersion relation.  This term is set to zero by \cite{GT} by construction to describe steady modes in a fully stable disk whereas Lau \& Bertin argue that out-of-phase terms are unlikely to be important for local stability.  As discussed in $\S$~\ref{sec:3dispersionrelation}, keeping these terms in gives us an effective dispersion relation with the ability to apply in a swing-amplification type scenario that is important for local stability in the regime of open spirals.  

A final difference from either \cite{laubertin} or \cite{GT} is that we proceed in 3D rather than integrating over the vertical direction to obtain a 2D version of the dispersion relation.  This conceptually changes the onset of conventional instability to 3D rather than 2D structures (and there are some quantitative changes, as described in $\S$~\ref{sec:conventionalQ}) and leaves vertical terms in the dispersion relation.  In practice, however, these vertical terms are negligible in the regime most relevant for structure growth, considerably simplifying the 3D dispersion relation (see $\S$~\ref{sec:conventionalQ}).


\begin{thebibliography}{}
\expandafter\ifx\csname natexlab\endcsname\relax\def\natexlab#1{#1}\fi

\bibitem[{{Aditya}(2023)}]{aditya23}
{Aditya}, K. 2023, \mnras, 522, 2543

\bibitem[{{Athanassoula}(1984)}]{athanassoula84}
{Athanassoula}, E. 1984, \physrep, 114, 319

\bibitem[{{Athanassoula}(1992)}]{athanassoula92}
---. 1992, \mnras, 259, 345

\bibitem[{{Baba} {et~al.}(2018){Baba}, {Kawata}, {Matsunaga}, {Grand}, \&
  {Hunt}}]{baba18}
{Baba}, J., {Kawata}, D., {Matsunaga}, N., {Grand}, R. J.~J., \& {Hunt}, J.
  A.~S. 2018, \apjl, 853, L23

\bibitem[{{Baba} {et~al.}(2016){Baba}, {Morokuma-Matsui}, {Miyamoto}, {Egusa},
  \& {Kuno}}]{baba16}
{Baba}, J., {Morokuma-Matsui}, K., {Miyamoto}, Y., {Egusa}, F., \& {Kuno}, N.
  2016, \mnras, 460, 2472

\bibitem[{{Baba} {et~al.}(2013){Baba}, {Saitoh}, \& {Wada}}]{baba13}
{Baba}, J., {Saitoh}, T.~R., \& {Wada}, K. 2013, \apj, 763, 46

\bibitem[{{Bertin} {et~al.}(1989){Bertin}, {Lin}, {Lowe}, \&
  {Thurstans}}]{bertin89}
{Bertin}, G., {Lin}, C.~C., {Lowe}, S.~A., \& {Thurstans}, R.~P. 1989, \apj,
  338, 78

\bibitem[{{Bertin} \& {Mark}(1978)}]{bertin78}
{Bertin}, G., \& {Mark}, J.~W.~K. 1978, \aap, 64, 389

\bibitem[{{Binney}(2020)}]{binney20swing}
{Binney}, J. 2020, \mnras, 496, 767

\bibitem[{{Binney} \& {Tremaine}(2008)}]{BT08}
{Binney}, J., \& {Tremaine}, S. 2008, {Galactic Dynamics: Second Edition}

\bibitem[{{Block} {et~al.}(2002){Block}, {Bournaud}, {Combes}, {Puerari}, \&
  {Buta}}]{block02}
{Block}, D.~L., {Bournaud}, F., {Combes}, F., {Puerari}, I., \& {Buta}, R.
  2002, \aap, 394, L35

\bibitem[{{Boissier} {et~al.}(2003){Boissier}, {Prantzos}, {Boselli}, \&
  {Gavazzi}}]{boissier03}
{Boissier}, S., {Prantzos}, N., {Boselli}, A., \& {Gavazzi}, G. 2003, \mnras,
  346, 1215

\bibitem[{{Bonaca} {et~al.}(2019){Bonaca}, {Hogg}, {Price-Whelan}, \&
  {Conroy}}]{bonaca19}
{Bonaca}, A., {Hogg}, D.~W., {Price-Whelan}, A.~M., \& {Conroy}, C. 2019, \apj,
  880, 38

\bibitem[{{Bournaud} {et~al.}(2011){Bournaud}, {Dekel}, {Teyssier}, {Cacciato},
  {Daddi}, {Juneau}, \& {Shankar}}]{bournaud11}
{Bournaud}, F., {Dekel}, A., {Teyssier}, R., {et~al.} 2011, \apjl, 741, L33

\bibitem[{{Cacciato} {et~al.}(2012{\natexlab{a}}){Cacciato}, {Dekel}, \&
  {Genel}}]{cacciato12}
{Cacciato}, M., {Dekel}, A., \& {Genel}, S. 2012{\natexlab{a}}, \mnras, 421,
  818

\bibitem[{{Cacciato} {et~al.}(2012{\natexlab{b}}){Cacciato}, {Dekel}, \&
  {Genel}}]{cacciato1}
---. 2012{\natexlab{b}}, \mnras, 421, 818

\bibitem[{{Chandrasekhar}(1951)}]{chandrasekhar51}
{Chandrasekhar}, S. 1951, Proceedings of the Royal Society of London Series A,
  210, 26

\bibitem[{{Chevance} {et~al.}(2022){Chevance}, {Kruijssen}, {Krumholz},
  {Groves}, {Keller}, {Hughes}, {Glover}, {Henshaw}, {Herrera}, {Kim}, {Leroy},
  {Pety}, {Razza}, {Rosolowsky}, {Schinnerer}, {Schruba}, {Barnes}, {Bigiel},
  {Blanc}, {Dale}, {Emsellem}, {Faesi}, {Grasha}, {Klessen}, {Kreckel}, {Liu},
  {Longmore}, {Meidt}, {Querejeta}, {Saito}, {Sun}, \& {Usero}}]{chevance}
{Chevance}, M., {Kruijssen}, J.~M.~D., {Krumholz}, M.~R., {et~al.} 2022,
  \mnras, 509, 272

\bibitem[{{Chiba}(2023)}]{chiba23}
{Chiba}, R. 2023, \mnras, 525, 3576

\bibitem[{{Chiba} \& {Sch{\"o}nrich}(2022)}]{chibaSch}
{Chiba}, R., \& {Sch{\"o}nrich}, R. 2022, \mnras, 513, 768

\bibitem[{{Combes} \& {Gerin}(1985)}]{combes85}
{Combes}, F., \& {Gerin}, M. 1985, \aap, 150, 327

\bibitem[{{De Rijcke} {et~al.}(2019){De Rijcke}, {Fouvry}, \&
  {Pichon}}]{derijcke}
{De Rijcke}, S., {Fouvry}, J.-B., \& {Pichon}, C. 2019, \mnras, 484, 3198

\bibitem[{{Dobbs} \& {Baba}(2014)}]{dobbsbaba}
{Dobbs}, C., \& {Baba}, J. 2014, \pasa, 31, e035

\bibitem[{{Dobbs} \& {Pringle}(2010)}]{dobbsspirals}
{Dobbs}, C.~L., \& {Pringle}, J.~E. 2010, \mnras, 409, 396

\bibitem[{{D'Onghia} {et~al.}(2013){D'Onghia}, {Vogelsberger}, \&
  {Hernquist}}]{donghia13}
{D'Onghia}, E., {Vogelsberger}, M., \& {Hernquist}, L. 2013, \apj, 766, 34

\bibitem[{{Dubinski} {et~al.}(2008){Dubinski}, {Gauthier}, {Widrow}, \&
  {Nickerson}}]{dubinski}
{Dubinski}, J., {Gauthier}, J.~R., {Widrow}, L., \& {Nickerson}, S. 2008, in
  Astronomical Society of the Pacific Conference Series, Vol. 396, Formation
  and Evolution of Galaxy Disks, ed. J.~G. {Funes} \& E.~M. {Corsini}, 321

\bibitem[{{Elmegreen}(1987)}]{elmegreen87}
{Elmegreen}, B.~G. 1987, \apj, 312, 626

\bibitem[{{Elmegreen}(1989)}]{elm89}
---. 1989, \apj, 344, 306

\bibitem[{{Elmegreen}(1994)}]{elmegreen94}
---. 1994, \apj, 433, 39

\bibitem[{{Elmegreen}(1995)}]{elmegreen95}
---. 1995, \mnras, 275, 944

\bibitem[{{Elmegreen}(2011{\natexlab{a}})}]{elm11}
---. 2011{\natexlab{a}}, \apj, 737, 10

\bibitem[{{Elmegreen}(2011{\natexlab{b}})}]{elmegreen11}
---. 2011{\natexlab{b}}, \apj, 737, 10

\bibitem[{{Elmegreen}(1980)}]{elmegreen80}
{Elmegreen}, D.~M. 1980, \apj, 242, 528

\bibitem[{{Elmegreen} \& {Elmegreen}(1982)}]{elmegreen82}
{Elmegreen}, D.~M., \& {Elmegreen}, B.~G. 1982, \mnras, 201, 1021

\bibitem[{{Ferguson} {et~al.}(2002){Ferguson}, {Irwin}, {Ibata}, {Lewis}, \&
  {Tanvir}}]{ferguson02}
{Ferguson}, A. M.~N., {Irwin}, M.~J., {Ibata}, R.~A., {Lewis}, G.~F., \&
  {Tanvir}, N.~R. 2002, \aj, 124, 1452

\bibitem[{{Forbes} {et~al.}(2012){Forbes}, {Krumholz}, \& {Burkert}}]{forbes12}
{Forbes}, J., {Krumholz}, M., \& {Burkert}, A. 2012, \apj, 754, 48

\bibitem[{{Forbes} {et~al.}(2014){Forbes}, {Krumholz}, {Burkert}, \&
  {Dekel}}]{forbes14}
{Forbes}, J.~C., {Krumholz}, M.~R., {Burkert}, A., \& {Dekel}, A. 2014, \mnras,
  438, 1552

\bibitem[{{Franx} {et~al.}(1994){Franx}, {van Gorkom}, \& {de Zeeuw}}]{franx}
{Franx}, M., {van Gorkom}, J.~H., \& {de Zeeuw}, T. 1994, \apj, 436, 642

\bibitem[{{Fuchs}(2001{\natexlab{a}})}]{fuchs3}
{Fuchs}, B. 2001{\natexlab{a}}, \mnras, 325, 1637

\bibitem[{{Fuchs}(2001{\natexlab{b}})}]{fuchs1}
---. 2001{\natexlab{b}}, \aap, 368, 107

\bibitem[{{Fujii} {et~al.}(2011){Fujii}, {Baba}, {Saitoh}, {Makino}, {Kokubo},
  \& {Wada}}]{fujii11}
{Fujii}, M.~S., {Baba}, J., {Saitoh}, T.~R., {et~al.} 2011, \apj, 730, 109

\bibitem[{{Fujii} {et~al.}(2018){Fujii}, {B{\'e}dorf}, {Baba}, \& {Portegies
  Zwart}}]{fujii18}
{Fujii}, M.~S., {B{\'e}dorf}, J., {Baba}, J., \& {Portegies Zwart}, S. 2018,
  \mnras, 477, 1451

\bibitem[{{Fujimoto}(1968)}]{fujimoto}
{Fujimoto}, M. 1968, \apj, 152, 391

\bibitem[{{Gammie}(2001)}]{gammie01}
{Gammie}, C.~F. 2001, \apj, 553, 174

\bibitem[{{Genel} {et~al.}(2012){Genel}, {Dekel}, \& {Cacciato}}]{genel12}
{Genel}, S., {Dekel}, A., \& {Cacciato}, M. 2012, \mnras, 425, 788

\bibitem[{{Genzel} {et~al.}(2013){Genzel}, {Tacconi}, {Kurk}, {Wuyts},
  {Combes}, {Freundlich}, {Bolatto}, {Cooper}, {Neri}, {Nordon}, {Bournaud},
  {Burkert}, {Comerford}, {Cox}, {Davis}, {F{\"o}rster Schreiber},
  {Garc{\'\i}a-Burillo}, {Gracia-Carpio}, {Lutz}, {Naab}, {Newman},
  {Saintonge}, {Shapiro Griffin}, {Shapley}, {Sternberg}, \&
  {Weiner}}]{genzel13}
{Genzel}, R., {Tacconi}, L.~J., {Kurk}, J., {et~al.} 2013, \apj, 773, 68

\bibitem[{{Ghosh} \& {Jog}(2018{\natexlab{a}})}]{ghoshjog1}
{Ghosh}, S., \& {Jog}, C.~J. 2018{\natexlab{a}}, \aap, 617, A47

\bibitem[{{Ghosh} \& {Jog}(2018{\natexlab{b}})}]{ghosh18}
---. 2018{\natexlab{b}}, \aap, 617, A47

\bibitem[{{Ghosh} \& {Jog}(2022)}]{ghoshjog2}
---. 2022, \aap, 658, A171

\bibitem[{{Goldreich} \& {Lynden-Bell}(1965{\natexlab{a}})}]{GLB}
{Goldreich}, P., \& {Lynden-Bell}, D. 1965{\natexlab{a}}, \mnras, 130, 97

\bibitem[{{Goldreich} \& {Lynden-Bell}(1965{\natexlab{b}})}]{GLBb}
---. 1965{\natexlab{b}}, \mnras, 130, 125

\bibitem[{{Goldreich} \& {Tremaine}(1978)}]{GT78}
{Goldreich}, P., \& {Tremaine}, S. 1978, \apj, 222, 850

\bibitem[{{Goldreich} \& {Tremaine}(1979)}]{GT}
---. 1979, \apj, 233, 857

\bibitem[{{Goz} {et~al.}(2015){Goz}, {Monaco}, {Murante}, \& {Curir}}]{goz15}
{Goz}, D., {Monaco}, P., {Murante}, G., \& {Curir}, A. 2015, \mnras, 447, 1774

\bibitem[{{Grand} {et~al.}(2015{\natexlab{a}}){Grand}, {Bovy}, {Kawata},
  {Hunt}, {Famaey}, {Siebert}, {Monari}, \& {Cropper}}]{grand15motions}
{Grand}, R. J.~J., {Bovy}, J., {Kawata}, D., {et~al.} 2015{\natexlab{a}},
  \mnras, 453, 1867

\bibitem[{{Grand} {et~al.}(2012){Grand}, {Kawata}, \& {Cropper}}]{grand12}
{Grand}, R. J.~J., {Kawata}, D., \& {Cropper}, M. 2012, \mnras, 426, 167

\bibitem[{{Grand} {et~al.}(2013){Grand}, {Kawata}, \& {Cropper}}]{grand13}
{Grand}, R.~J.~J., {Kawata}, D., \& {Cropper}, M. 2013, \aap, 553, A77

\bibitem[{{Grand} {et~al.}(2015{\natexlab{b}}){Grand}, {Kawata}, \&
  {Cropper}}]{grand15}
{Grand}, R. J.~J., {Kawata}, D., \& {Cropper}, M. 2015{\natexlab{b}}, \mnras,
  447, 4018

\bibitem[{{Grand} {et~al.}(2016){Grand}, {Springel}, {Kawata}, {Minchev},
  {S{\'a}nchez-Bl{\'a}zquez}, {G{\'o}mez}, {Marinacci}, {Pakmor}, \&
  {Campbell}}]{grand16}
{Grand}, R. J.~J., {Springel}, V., {Kawata}, D., {et~al.} 2016, \mnras, 460,
  L94

\bibitem[{{Grand} {et~al.}(2017){Grand}, {G{\'o}mez}, {Marinacci}, {Pakmor},
  {Springel}, {Campbell}, {Frenk}, {Jenkins}, \& {White}}]{grand17}
{Grand}, R. J.~J., {G{\'o}mez}, F.~A., {Marinacci}, F., {et~al.} 2017, \mnras,
  467, 179

\bibitem[{{Grosb{\o}l} {et~al.}(2004){Grosb{\o}l}, {Patsis}, \&
  {Pompei}}]{grosbol04}
{Grosb{\o}l}, P., {Patsis}, P.~A., \& {Pompei}, E. 2004, \aap, 423, 849

\bibitem[{{Hamilton}(2023)}]{hamilton}
{Hamilton}, C. 2023, arXiv e-prints, arXiv:2302.06602

\bibitem[{{Hart} {et~al.}(2018){Hart}, {Bamford}, {Keel}, {Kruk}, {Masters},
  {Simmons}, \& {Smethurst}}]{hart18}
{Hart}, R.~E., {Bamford}, S.~P., {Keel}, W.~C., {et~al.} 2018, \mnras, 478, 932

\bibitem[{{Hart} {et~al.}(2017){Hart}, {Bamford}, {Hayes}, {Cardamone}, {Keel},
  {Kruk}, {Lintott}, {Masters}, {Simmons}, \& {Smethurst}}]{hart17}
{Hart}, R.~E., {Bamford}, S.~P., {Hayes}, W.~B., {et~al.} 2017, \mnras, 472,
  2263

\bibitem[{{Hopkins} \& {Quataert}(2011)}]{hopkins11}
{Hopkins}, P.~F., \& {Quataert}, E. 2011, \mnras, 415, 1027

\bibitem[{{Hunt} {et~al.}(2018){Hunt}, {Hong}, {Bovy}, {Kawata}, \&
  {Grand}}]{hunt18}
{Hunt}, J. A.~S., {Hong}, J., {Bovy}, J., {Kawata}, D., \& {Grand}, R. J.~J.
  2018, \mnras, 481, 3794

\bibitem[{{Jog}(1992)}]{jogswing}
{Jog}, C.~J. 1992, \apj, 390, 378

\bibitem[{{Jog}(1996)}]{jog96}
---. 1996, \mnras, 278, 209

\bibitem[{{Jog} \& {Solomon}(1984)}]{jogsolomon}
{Jog}, C.~J., \& {Solomon}, P.~M. 1984, \apj, 276, 114

\bibitem[{{Julian} \& {Toomre}(1966)}]{JT}
{Julian}, W.~H., \& {Toomre}, A. 1966, \apj, 146, 810

\bibitem[{{Kalnajs}(1965)}]{kalnajs65}
{Kalnajs}, A.~J. 1965, PhD thesis, Harvard University, Massachusetts

\bibitem[{{Kalnajs}(1972)}]{kalnajs}
---. 1972, \apj, 175, 63

\bibitem[{{Kawata} {et~al.}(2014){Kawata}, {Hunt}, {Grand}, {Pasetto}, \&
  {Cropper}}]{kawata}
{Kawata}, D., {Hunt}, J. A.~S., {Grand}, R. J.~J., {Pasetto}, S., \& {Cropper},
  M. 2014, \mnras, 443, 2757

\bibitem[{{Kendall} {et~al.}(2015){Kendall}, {Clarke}, \&
  {Kennicutt}}]{kendall15}
{Kendall}, S., {Clarke}, C., \& {Kennicutt}, R.~C. 2015, \mnras, 446, 4155

\bibitem[{{Kennicutt}(1981)}]{kennicutt81}
{Kennicutt}, R.~C., J. 1981, \aj, 86, 1847

\bibitem[{{Khoperskov} \& {Khrapov}(2018)}]{Khoperskov18}
{Khoperskov}, S.~A., \& {Khrapov}, S.~S. 2018, \aap, 609, A104

\bibitem[{{Kim} {et~al.}(2022){Kim}, {Chevance}, {Kruijssen}, {Leroy},
  {Schruba}, {Barnes}, {Bigiel}, {Blanc}, {Cao}, {Congiu}, {Dale}, {Faesi},
  {Glover}, {Grasha}, {Groves}, {Hughes}, {Klessen}, {Kreckel}, {McElroy},
  {Pan}, {Pety}, {Querejeta}, {Razza}, {Rosolowsky}, {Saito}, {Schinnerer},
  {Sun}, {Tomi{\v{c}}i{\'c}}, {Usero}, \& {Williams}}]{kim22}
{Kim}, J., {Chevance}, M., {Kruijssen}, J.~M.~D., {et~al.} 2022, \mnras, 516,
  3006

\bibitem[{{Kim} {et~al.}(2021){Kim}, {Ostriker}, \& {Filippova}}]{kim21fFB}
{Kim}, J.-G., {Ostriker}, E.~C., \& {Filippova}, N. 2021, \apj, 911, 128

\bibitem[{{Kim} {et~al.}(2020{\natexlab{a}}){Kim}, {Kim}, \&
  {Ostriker}}]{kko20}
{Kim}, W.-T., {Kim}, C.-G., \& {Ostriker}, E.~C. 2020{\natexlab{a}}, \apj, 898,
  35

\bibitem[{{Kim} {et~al.}(2020{\natexlab{b}}){Kim}, {Kim}, \&
  {Ostriker}}]{kim20}
---. 2020{\natexlab{b}}, \apj, 898, 35

\bibitem[{{Kim} \& {Ostriker}(2000)}]{ko2000}
{Kim}, W.-T., \& {Ostriker}, E.~C. 2000, \apj, 540, 372

\bibitem[{{Kim} \& {Ostriker}(2001{\natexlab{a}})}]{ko01Qthresh}
---. 2001{\natexlab{a}}, \apj, 559, 70

\bibitem[{{Kim} \& {Ostriker}(2001{\natexlab{b}})}]{kimOstriker01}
---. 2001{\natexlab{b}}, \apj, 559, 70

\bibitem[{{Kim} \& {Ostriker}(2002)}]{ko02spurs}
---. 2002, \apj, 570, 132

\bibitem[{{Kim} \& {Ostriker}(2006{\natexlab{a}})}]{ko06}
---. 2006{\natexlab{a}}, \apj, 646, 213

\bibitem[{{Kim} \& {Ostriker}(2006{\natexlab{b}})}]{kim06}
---. 2006{\natexlab{b}}, \apj, 646, 213

\bibitem[{{Kim} {et~al.}(2012){Kim}, {Seo}, \& {Kim}}]{kim12bars}
{Kim}, W.-T., {Seo}, W.-Y., \& {Kim}, Y. 2012, \apj, 758, 14

\bibitem[{{Koyama} \& {Ostriker}(2009)}]{koyama}
{Koyama}, H., \& {Ostriker}, E.~C. 2009, \apj, 693, 1346

\bibitem[{{Krumholz} \& {Burkert}(2010)}]{krumBurk}
{Krumholz}, M., \& {Burkert}, A. 2010, \apj, 724, 895

\bibitem[{{Kumamoto} \& {Noguchi}(2016)}]{kumamoto16}
{Kumamoto}, J., \& {Noguchi}, M. 2016, \apj, 822, 110

\bibitem[{{Kwak} {et~al.}(2019){Kwak}, {Kim}, {Rey}, \& {Quinn}}]{kwak19}
{Kwak}, S., {Kim}, W.-T., {Rey}, S.-C., \& {Quinn}, T.~R. 2019, \apj, 887, 139

\bibitem[{{La Vigne} {et~al.}(2006{\natexlab{a}}){La Vigne}, {Vogel}, \&
  {Ostriker}}]{lavigne}
{La Vigne}, M.~A., {Vogel}, S.~N., \& {Ostriker}, E.~C. 2006{\natexlab{a}},
  \apj, 650, 818

\bibitem[{{La Vigne} {et~al.}(2006{\natexlab{b}}){La Vigne}, {Vogel}, \&
  {Ostriker}}]{lavigne06}
---. 2006{\natexlab{b}}, \apj, 650, 818

\bibitem[{{Lau} \& {Bertin}(1978)}]{laubertin}
{Lau}, Y.~Y., \& {Bertin}, G. 1978, \apj, 226, 508

\bibitem[{{Lau} {et~al.}(1976){Lau}, {Lin}, \& {Mark}}]{laulinmark}
{Lau}, Y.~Y., {Lin}, C.~C., \& {Mark}, J. W.~K. 1976, Proceedings of the
  National Academy of Science, 73, 1379

\bibitem[{{Leroy} {et~al.}(2008){Leroy}, {Walter}, {Brinks}, {Bigiel}, {de
  Blok}, {Madore}, \& {Thornley}}]{leroy08}
{Leroy}, A.~K., {Walter}, F., {Brinks}, E., {et~al.} 2008, \aj, 136, 2782

\bibitem[{{Leroy} {et~al.}(2023){Leroy}, {Sandstrom}, {Rosolowsky}, {Belfiore},
  {Bolatto}, {Cao}, {Koch}, {Schinnerer}, {Barnes}, {Be{\v{s}}li{\'c}},
  {Bigiel}, {Blanc}, {Chastenet}, {Chen}, {Chevance}, {Chown}, {Congiu},
  {Dale}, {Egorov}, {Emsellem}, {Eibensteiner}, {Faesi}, {Glover}, {Grasha},
  {Groves}, {Hassani}, {Henshaw}, {Hughes}, {Jim{\'e}nez-Donaire}, {Kim},
  {Klessen}, {Kreckel}, {Kruijssen}, {Larson}, {Lee}, {Levy}, {Liu}, {Lopez},
  {Meidt}, {Murphy}, {Neumann}, {Pessa}, {Pety}, {Saito}, {Sardone}, {Sun},
  {Thilker}, {Usero}, {Watkins}, {Whitcomb}, \& {Williams}}]{leroy23}
{Leroy}, A.~K., {Sandstrom}, K., {Rosolowsky}, E., {et~al.} 2023, \apjl, 944,
  L9

\bibitem[{{Li} {et~al.}(2005){Li}, {Mac Low}, \& {Klessen}}]{li05}
{Li}, Y., {Mac Low}, M.-M., \& {Klessen}, R.~S. 2005, \apj, 626, 823

\bibitem[{{Lin} \& {Shu}(1964)}]{linshu}
{Lin}, C.~C., \& {Shu}, F.~H. 1964, \apj, 140, 646

\bibitem[{{Lin} \& {Shu}(1966)}]{linshu66}
---. 1966, Proceedings of the National Academy of Science, 55, 229

\bibitem[{{Lingard} {et~al.}(2021){Lingard}, {Masters}, {Krawczyk}, {Lintott},
  {Kruk}, {Simmons}, {Keel}, {Nichol}, \& {Baeten}}]{lingard21}
{Lingard}, T., {Masters}, K.~L., {Krawczyk}, C., {et~al.} 2021, \mnras, 504,
  3364

\bibitem[{{Lynden-Bell} \& {Kalnajs}(1972)}]{LBK}
{Lynden-Bell}, D., \& {Kalnajs}, A.~J. 1972, \mnras, 157, 1

\bibitem[{{Mac Low} {et~al.}(2017){Mac Low}, {Burkert}, \&
  {Ib{\'a}{\~n}ez-Mej{\'\i}a}}]{maclow17}
{Mac Low}, M.-M., {Burkert}, A., \& {Ib{\'a}{\~n}ez-Mej{\'\i}a}, J.~C. 2017,
  \apjl, 847, L10

\bibitem[{{Mandowara} {et~al.}(2022){Mandowara}, {Sormani}, {Sobacchi}, \&
  {Klessen}}]{mandowara}
{Mandowara}, Y., {Sormani}, M.~C., {Sobacchi}, E., \& {Klessen}, R.~S. 2022,
  \mnras, 513, 5052

\bibitem[{{Mark}(1974)}]{mark74}
{Mark}, J.~W.~K. 1974, \apj, 193, 539

\bibitem[{{Mark}(1976)}]{mark76}
---. 1976, \apj, 206, 418

\bibitem[{{Mark}(1977)}]{mark77}
---. 1977, \apj, 212, 645

\bibitem[{{Martin} \& {Kennicutt}(2001)}]{martin01}
{Martin}, C.~L., \& {Kennicutt}, Robert~C., J. 2001, \apj, 555, 301

\bibitem[{{Maseda} {et~al.}(2014){Maseda}, {van der Wel}, {Rix}, {da Cunha},
  {Pacifici}, {Momcheva}, {Brammer}, {Meidt}, {Franx}, {van Dokkum},
  {Fumagalli}, {Bell}, {Ferguson}, {F{\"o}rster-Schreiber}, {Koekemoer}, {Koo},
  {Lundgren}, {Marchesini}, {Nelson}, {Patel}, {Skelton}, {Straughn}, {Trump},
  \& {Whitaker}}]{maseda14}
{Maseda}, M.~V., {van der Wel}, A., {Rix}, H.-W., {et~al.} 2014, \apj, 791, 17

\bibitem[{{Masters} {et~al.}(2019){Masters}, {Lintott}, {Hart}, {Kruk},
  {Smethurst}, {Casteels}, {Keel}, {Simmons}, {Stanescu}, {Tate}, \&
  {Tomi}}]{masters18}
{Masters}, K.~L., {Lintott}, C.~J., {Hart}, R.~E., {et~al.} 2019, \mnras, 487,
  1808

\bibitem[{{Meidt}(2022)}]{meidt22}
{Meidt}, S.~E. 2022, \apj, 937, 88

\bibitem[{{Meidt} {et~al.}(2018){Meidt}, {Leroy}, {Rosolowsky}, {Kruijssen},
  {Schinnerer}, {Schruba}, {Pety}, {Blanc}, {Bigiel}, {Chevance}, {Hughes},
  {Querejeta}, \& {Usero}}]{meidt18}
{Meidt}, S.~E., {Leroy}, A.~K., {Rosolowsky}, E., {et~al.} 2018, \apj, 854, 100

\bibitem[{{Meidt} {et~al.}(2020){Meidt}, {Glover}, {Kruijssen}, {Leroy},
  {Rosolowsky}, {Hughes}, {Schinnerer}, {Schruba}, {Usero}, {Bigiel}, {Blanc},
  {Chevance}, {Pety}, {Querejeta}, \& {Utomo}}]{meidt20}
{Meidt}, S.~E., {Glover}, S. C.~O., {Kruijssen}, J.~M.~D., {et~al.} 2020, \apj,
  892, 73

\bibitem[{{Meidt} {et~al.}(2023{\natexlab{a}}){Meidt}, {Rosolowsky}, {Sun},
  {Koch}, {Klessen}, {Leroy}, {Schinnerer}, {Barnes}, {Glover}, {Lee}, {van der
  Wel}, {Watkins}, {Williams}, {Bigiel}, {Boquien}, {Blanc}, {Cao}, {Chevance},
  {Dale}, {Egorov}, {Emsellem}, {Grasha}, {Henshaw}, {Kruijssen}, {Larson},
  {Liu}, {Murphy}, {Pety}, {Querejeta}, {Saito}, {Sandstrom}, {Smith},
  {Sormani}, \& {Thilker}}]{williams23}
{Meidt}, S.~E., {Rosolowsky}, E., {Sun}, J., {et~al.} 2023{\natexlab{a}},
  \apjl, 944, L18

\bibitem[{{Meidt} {et~al.}(2023{\natexlab{b}}){Meidt}, {Rosolowsky}, {Sun},
  {Koch}, {Klessen}, {Leroy}, {Schinnerer}, {Barnes}, {Glover}, {Lee}, {van der
  Wel}, {Watkins}, {Williams}, {Bigiel}, {Boquien}, {Blanc}, {Cao}, {Chevance},
  {Dale}, {Egorov}, {Emsellem}, {Grasha}, {Henshaw}, {Kruijssen}, {Larson},
  {Liu}, {Murphy}, {Pety}, {Querejeta}, {Saito}, {Sandstrom}, {Smith},
  {Sormani}, \& {Thilker}}]{meidt23}
---. 2023{\natexlab{b}}, \apjl, 944, L18

\bibitem[{{Merrifield} {et~al.}(2006){Merrifield}, {Rand}, \&
  {Meidt}}]{merrifield06}
{Merrifield}, M.~R., {Rand}, R.~J., \& {Meidt}, S.~E. 2006, \mnras, 366, L17

\bibitem[{{Michikoshi} \& {Kokubo}(2016)}]{michikoshi16}
{Michikoshi}, S., \& {Kokubo}, E. 2016, \apj, 821, 35

\bibitem[{{Michikoshi} \& {Kokubo}(2018)}]{michikoshi18}
---. 2018, \mnras, 481, 185

\bibitem[{{Minchev} {et~al.}(2012){Minchev}, {Famaey}, {Quillen}, {Di Matteo},
  {Combes}, {Vlaji{\'c}}, {Erwin}, \& {Bland-Hawthorn}}]{minchev12}
{Minchev}, I., {Famaey}, B., {Quillen}, A.~C., {et~al.} 2012, \aap, 548, A126

\bibitem[{{Mollitor} {et~al.}(2015){Mollitor}, {Nezri}, \&
  {Teyssier}}]{mollitor15}
{Mollitor}, P., {Nezri}, E., \& {Teyssier}, R. 2015, \mnras, 447, 1353

\bibitem[{{Moore} {et~al.}(1999){Moore}, {Ghigna}, {Governato}, {Lake},
  {Quinn}, {Stadel}, \& {Tozzi}}]{moore99}
{Moore}, B., {Ghigna}, S., {Governato}, F., {et~al.} 1999, \apjl, 524, L19

\bibitem[{{Morozov}(1985)}]{morozov}
{Morozov}, A.~G. 1985, \sovast, 29, 120

\bibitem[{{Mutlu-Pakdil} {et~al.}(2018){Mutlu-Pakdil}, {Seigar}, {Hewitt},
  {Treuthardt}, {Berrier}, \& {Koval}}]{mutlu18}
{Mutlu-Pakdil}, B., {Seigar}, M.~S., {Hewitt}, I.~B., {et~al.} 2018, \mnras,
  474, 2594

\bibitem[{{Nelson} {et~al.}(2012){Nelson}, {D'Onghia}, \&
  {Hernquist}}]{nelson12}
{Nelson}, D., {D'Onghia}, E., \& {Hernquist}, L. 2012, in Astronomical Society
  of the Pacific Conference Series, Vol. 453, Advances in Computational
  Astrophysics: Methods, Tools, and Outcome, ed. R.~{Capuzzo-Dolcetta},
  M.~{Limongi}, \& A.~{Tornamb{\`e}}, 369

\bibitem[{{Pettitt} {et~al.}(2016){Pettitt}, {Tasker}, \&
  {Wadsley}}]{pettitt16}
{Pettitt}, A.~R., {Tasker}, E.~J., \& {Wadsley}, J.~W. 2016, \mnras, 458, 3990

\bibitem[{{Pettitt} {et~al.}(2017){Pettitt}, {Tasker}, {Wadsley}, {Keller}, \&
  {Benincasa}}]{pettitt17}
{Pettitt}, A.~R., {Tasker}, E.~J., {Wadsley}, J.~W., {Keller}, B.~W., \&
  {Benincasa}, S.~M. 2017, \mnras, 468, 4189

\bibitem[{{Pfenniger}(1990)}]{pfenniger}
{Pfenniger}, D. 1990, \aap, 230, 55

\bibitem[{{Pringle} \& {Dobbs}(2019)}]{pringle19}
{Pringle}, J.~E., \& {Dobbs}, C.~L. 2019, \mnras, 490, 1470

\bibitem[{{Rafikov}(2001)}]{rafikov}
{Rafikov}, R.~R. 2001, \mnras, 323, 445

\bibitem[{{Roberts} \& {Stewart}(1987)}]{robertsstewart}
{Roberts}, William~W., J., \& {Stewart}, G.~R. 1987, \apj, 314, 10

\bibitem[{{Roberts}(1969)}]{roberts69}
{Roberts}, W.~W. 1969, \apj, 158, 123

\bibitem[{{Romeo}(1992)}]{romeo92}
{Romeo}, A.~B. 1992, \mnras, 256, 307

\bibitem[{{Romeo} {et~al.}(2010){Romeo}, {Burkert}, \& {Agertz}}]{romeo10}
{Romeo}, A.~B., {Burkert}, A., \& {Agertz}, O. 2010, \mnras, 407, 1223

\bibitem[{{Romeo} \& {Falstad}(2013)}]{romeo13}
{Romeo}, A.~B., \& {Falstad}, N. 2013, \mnras, 433, 1389

\bibitem[{{Romeo} \& {Wiegert}(2011)}]{romeo11}
{Romeo}, A.~B., \& {Wiegert}, J. 2011, \mnras, 416, 1191

\bibitem[{{Ro{\v{s}}kar} {et~al.}(2012){Ro{\v{s}}kar}, {Debattista}, {Quinn},
  \& {Wadsley}}]{roskar12}
{Ro{\v{s}}kar}, R., {Debattista}, V.~P., {Quinn}, T.~R., \& {Wadsley}, J. 2012,
  \mnras, 426, 2089

\bibitem[{{Sancisi} {et~al.}(2008){Sancisi}, {Fraternali}, {Oosterloo}, \& {van
  der Hulst}}]{sancisi08}
{Sancisi}, R., {Fraternali}, F., {Oosterloo}, T., \& {van der Hulst}, T. 2008,
  \aapr, 15, 189

\bibitem[{{Sanderson} {et~al.}(2020){Sanderson}, {Wetzel}, {Loebman}, {Sharma},
  {Hopkins}, {Garrison-Kimmel}, {Faucher-Gigu{\`e}re}, {Kere{\v{s}}}, \&
  {Quataert}}]{sanderson20}
{Sanderson}, R.~E., {Wetzel}, A., {Loebman}, S., {et~al.} 2020, \apjs, 246, 6

\bibitem[{{Sandstrom} {et~al.}(2023){Sandstrom}, {Koch}, {Leroy}, {Rosolowsky},
  {Emsellem}, {Smith}, {Egorov}, {Williams}, {Larson}, {Lee}, {Schinnerer},
  {Thilker}, {Barnes}, {Belfiore}, {Bigiel}, {Blanc}, {Bolatto}, {Boquien},
  {Cao}, {Chastenet}, {Chevance}, {Chiang}, {Dale}, {Faesi}, {Glover},
  {Grasha}, {Groves}, {Hassani}, {Henshaw}, {Hughes}, {Kim}, {Klessen},
  {Kreckel}, {Kruijssen}, {Lopez}, {Liu}, {Meidt}, {Murphy}, {Pan},
  {Querejeta}, {Saito}, {Sardone}, {Sormani}, {Sutter}, {Usero}, \&
  {Watkins}}]{sandstrom23}
{Sandstrom}, K.~M., {Koch}, E.~W., {Leroy}, A.~K., {et~al.} 2023, \apjl, 944,
  L8

\bibitem[{{Seigar} \& {James}(1998)}]{seigar98}
{Seigar}, M.~S., \& {James}, P.~A. 1998, \mnras, 299, 685

\bibitem[{{Sellwood}(2011{\natexlab{a}})}]{sellwoodlifetimes}
{Sellwood}, J.~A. 2011{\natexlab{a}}, \mnras, 410, 1637

\bibitem[{{Sellwood}(2011{\natexlab{b}})}]{sellwood11}
---. 2011{\natexlab{b}}, \mnras, 410, 1637

\bibitem[{{Sellwood} \& {Binney}(2002)}]{sellwoodbinney}
{Sellwood}, J.~A., \& {Binney}, J.~J. 2002, \mnras, 336, 785

\bibitem[{{Sellwood} \& {Carlberg}(2014)}]{SC14}
{Sellwood}, J.~A., \& {Carlberg}, R.~G. 2014, \apj, 785, 137

\bibitem[{{Sellwood} \& {Carlberg}(2019{\natexlab{a}})}]{SC}
---. 2019{\natexlab{a}}, \mnras, 489, 116

\bibitem[{{Sellwood} \& {Carlberg}(2019{\natexlab{b}})}]{sellwood19}
---. 2019{\natexlab{b}}, \mnras, 489, 116

\bibitem[{{Sellwood} \& {Carlberg}(2021)}]{SCnonlin}
---. 2021, \mnras, 500, 5043

\bibitem[{{Sellwood} \& {Carlberg}(2022)}]{SCsat}
---. 2022, \mnras, 517, 2610

\bibitem[{{Sellwood} \& {Masters}(2022)}]{sellwoodmasters}
{Sellwood}, J.~A., \& {Masters}, K.~L. 2022, \araa, 60, arXiv:2110.05615

\bibitem[{{Shetty} \& {Ostriker}(2006{\natexlab{a}})}]{shetty}
{Shetty}, R., \& {Ostriker}, E.~C. 2006{\natexlab{a}}, \apj, 647, 997

\bibitem[{{Shetty} \& {Ostriker}(2006{\natexlab{b}})}]{shetty06}
---. 2006{\natexlab{b}}, \apj, 647, 997

\bibitem[{{Snyder} {et~al.}(2015){Snyder}, {Torrey}, {Lotz}, {Genel},
  {McBride}, {Vogelsberger}, {Pillepich}, {Nelson}, {Sales}, {Sijacki},
  {Hernquist}, \& {Springel}}]{snyder15}
{Snyder}, G.~F., {Torrey}, P., {Lotz}, J.~M., {et~al.} 2015, \mnras, 454, 1886

\bibitem[{{Sormani} {et~al.}(2017){Sormani}, {Sobacchi}, {Shore}, {Tre{\ss}},
  \& {Klessen}}]{sormani}
{Sormani}, M.~C., {Sobacchi}, E., {Shore}, S.~N., {Tre{\ss}}, R.~G., \&
  {Klessen}, R.~S. 2017, \mnras, 471, 2932

\bibitem[{{Stott} {et~al.}(2016){Stott}, {Swinbank}, {Johnson}, {Tiley},
  {Magdis}, {Bower}, {Bunker}, {Bureau}, {Harrison}, {Jarvis}, {Sharples},
  {Smail}, {Sobral}, {Best}, \& {Cirasuolo}}]{stott16}
{Stott}, J.~P., {Swinbank}, A.~M., {Johnson}, H.~L., {et~al.} 2016, \mnras,
  457, 1888

\bibitem[{{Thilker} {et~al.}(2023){Thilker}, {Lee}, {Deger}, {Barnes},
  {Bigiel}, {Boquien}, {Cao}, {Chevance}, {Dale}, {Egorov}, {Glover}, {Grasha},
  {Henshaw}, {Klessen}, {Koch}, {Kruijssen}, {Leroy}, {Lessing}, {Meidt},
  {Pinna}, {Querejeta}, {Rosolowsky}, {Sandstrom}, {Schinnerer}, {Smith},
  {Watkins}, {Williams}, {Anand}, {Belfiore}, {Blanc}, {Chandar}, {Congiu},
  {Emsellem}, {Groves}, {Kreckel}, {Larson}, {Liu}, {Pessa}, \&
  {Whitmore}}]{thilker23}
{Thilker}, D.~A., {Lee}, J.~C., {Deger}, S., {et~al.} 2023, \apjl, 944, L13

\bibitem[{{Toomre}(1964)}]{toomre64}
{Toomre}, A. 1964, \apj, 139, 1217

\bibitem[{{Toomre}(1969)}]{toomre69}
---. 1969, \apj, 158, 899

\bibitem[{{Toomre}(1977)}]{toomre77}
---. 1977, \araa, 15, 437

\bibitem[{{Toomre}(1981)}]{toomre81}
{Toomre}, A. 1981, in Structure and Evolution of Normal Galaxies, ed. S.~M.
  {Fall} \& D.~{Lynden-Bell}, 111--136

\bibitem[{{Tremaine} \& {Weinberg}(1984)}]{TW84}
{Tremaine}, S., \& {Weinberg}, M.~D. 1984, \mnras, 209, 729

\bibitem[{{{\"U}bler} {et~al.}(2019){{\"U}bler}, {Genzel}, {Wisnioski},
  {F{\"o}rster Schreiber}, {Shimizu}, {Price}, {Tacconi}, {Belli}, {Wilman},
  {Fossati}, {Mendel}, {Davies}, {Beifiori}, {Bender}, {Brammer}, {Burkert},
  {Chan}, {Davies}, {Fabricius}, {Galametz}, {Herrera-Camus}, {Lang}, {Lutz},
  {Momcheva}, {Naab}, {Nelson}, {Saglia}, {Tadaki}, {van Dokkum}, \&
  {Wuyts}}]{ubler19}
{{\"U}bler}, H., {Genzel}, R., {Wisnioski}, E., {et~al.} 2019, \apj, 880, 48

\bibitem[{{Villanueva} {et~al.}(2021){Villanueva}, {Bolatto}, {Vogel}, {Levy},
  {S{\'a}nchez}, {Barrera-Ballesteros}, {Wong}, {Rosolowsky}, {Colombo},
  {Rubio}, {Cao}, {Kalinova}, {Leroy}, {Utomo}, {Herrera-Camus}, {Blitz}, \&
  {Luo}}]{villanueva21}
{Villanueva}, V., {Bolatto}, A., {Vogel}, S., {et~al.} 2021, \apj, 923, 60

\bibitem[{{Wada}(2008)}]{wada08}
{Wada}, K. 2008, \apj, 675, 188

\bibitem[{{Wada} {et~al.}(2011){Wada}, {Baba}, \& {Saitoh}}]{wadaspirals}
{Wada}, K., {Baba}, J., \& {Saitoh}, T.~R. 2011, \apj, 735, 1

\bibitem[{{Wada} \& {Koda}(2004)}]{wadakoda}
{Wada}, K., \& {Koda}, J. 2004, \mnras, 349, 270

\bibitem[{{Watkins} {et~al.}(2023){Watkins}, {Barnes}, {Henny}, {Kim},
  {Kreckel}, {Meidt}, {Klessen}, {Glover}, {Williams}, {Keller}, {Leroy},
  {Rosolowsky}, {Lee}, {Anand}, {Belfiore}, {Bigiel}, {Blanc}, {Boquien},
  {Cao}, {Chandar}, {Chen}, {Chevance}, {Congiu}, {Dale}, {Deger}, {Egorov},
  {Emsellem}, {Faesi}, {Grasha}, {Groves}, {Hassani}, {Henshaw}, {Herrera},
  {Hughes}, {Jeffreson}, {Jim{\'e}nez-Donaire}, {Koch}, {Kruijssen}, {Larson},
  {Liu}, {Lopez}, {Pessa}, {Pety}, {Querejeta}, {Saito}, {Sandstrom},
  {Scheuermann}, {Schinnerer}, {Sormani}, {Stuber}, {Thilker}, {Usero}, \&
  {Whitmore}}]{watkins23}
{Watkins}, E.~J., {Barnes}, A.~T., {Henny}, K., {et~al.} 2023, \apjl, 944, L24

\bibitem[{{Weinberg}(2004)}]{weinberg04}
{Weinberg}, M.~D. 2004, arXiv e-prints, astro

\bibitem[{{Weinberg} \& {Katz}(2007{\natexlab{a}})}]{weinbergKatz07a}
{Weinberg}, M.~D., \& {Katz}, N. 2007{\natexlab{a}}, \mnras, 375, 425

\bibitem[{{Weinberg} \& {Katz}(2007{\natexlab{b}})}]{weinbergKatz07b}
---. 2007{\natexlab{b}}, \mnras, 375, 460

\bibitem[{{Westfall} {et~al.}(2014){Westfall}, {Andersen}, {Bershady},
  {Martinsson}, {Swaters}, \& {Verheijen}}]{westfall14}
{Westfall}, K.~B., {Andersen}, D.~R., {Bershady}, M.~A., {et~al.} 2014, \apj,
  785, 43

\bibitem[{{Yu} \& {Ho}(2019)}]{yu19}
{Yu}, S.-Y., \& {Ho}, L.~C. 2019, \apj, 871, 194

\end{thebibliography}
 \end{document}